\documentclass[
    journal,
    final,
    10pt,
    a4paper
    ]{IEEEtran}
    
\usepackage{graphicx} 
\graphicspath{{../figures/}}
\usepackage[tight,footnotesize]{subfigure} 
\usepackage{multirow}
\usepackage{array} 
\usepackage{cite} 
\usepackage{color}
\usepackage{url} 
\usepackage{amssymb} 
\usepackage{amsmath}
\usepackage{breqn} 
\usepackage{dsfont} 
\usepackage{enumerate}
\usepackage{mathrsfs} 
\usepackage{mathtools} 
\usepackage[T1]{fontenc}

\def\E{\,\mathds{E}\,}   \def\a{{\bf a}}
  \def\x{{\bf x}} 
 \def\h{{\bf h}} \def\e{{\bf e}} \def\f{{\bf f}}
 \def\v{{\bf v}} \def\w{{\bf w}} 
  \def\y{{\bf y}} \def\p{{\bf p}}
  \def\I{{\bf I}} \def\P{{\bf P}}
\def\Q{{\bf Q}} \def\A{{\bf A}} \def\B{{\bf B}} \def\C{{\bf C}}
  \def\J{{\bf J}} 
\def\G{{\bf G}}   \def\R{\mathds{R}}
\def\W{{\bf W}} \def\s{{\bf s}} \def\D{{\bf D}}

\def\1{\mathbf{1}}

\newtheorem{theorem}{Theorem}[section] \newtheorem{lemma}[theorem]{\bf
  Lemma} \newtheorem{proposition}[theorem]{\bf Proposition}

\newenvironment{remark}[1][Remark]{\begin{trivlist}
\item[\hskip \labelsep {\bfseries #1}]}{\end{trivlist}}

\newcommand{\qed}{\nobreak \ifvmode \relax \else
  \ifdim\lastskip<1.5em \hskip-\lastskip \hskip1.5em plus0em
  minus0.5em \fi \nobreak \vrule height0.75em width0.5em
  depth0.25em\fi}

\begin{document}
%
\title{Mobile Node Localization via Pareto Optimization: Algorithm and Fundamental Performance Limitations}
%
%
%

\author{Alessio De Angelis,~\IEEEmembership{Member,~IEEE}, Carlo
  Fischione,~\IEEEmembership{Member,~IEEE} 
  \thanks{ Alessio De Angelis is with the Department of Engineering, 
    University of Perugia,
    Italy.  E-mail: alessio.deangelis{@}unipg.it.
    Carlo Fischione is with ACCESS Linnaeus Centre, Department of
    Electrical Engineering, KTH Royal Institute of Technology,
    Stockholm, Sweden.  E-mail: carlofi{@}kth.se}}%
\IEEEpubid{0000--0000/00\$00.00~\copyright~20xx IEEE}

\maketitle
\thispagestyle{empty}

\vspace{-2cm}

\begin{abstract}
Accurate estimation of the position of network nodes is essential, e.g., in  localization, geographic routing, and
vehicular networks. Unfortunately, typical positioning techniques
based on ranging or on velocity and angular measurements are
inherently limited. To overcome the limitations of specific
positioning techniques, the fusion of multiple and heterogeneous sensor information
is an appealing strategy. In this paper, we investigate the
fundamental performance of linear fusion of multiple measurements
of the position of mobile nodes, and propose a new distributed
recursive position estimator. The Cram\'er-Rao lower bounds for the
parametric and a-posteriori cases are investigated. The proposed
estimator combines information coming from ranging, speed, and angular
measurements, which is jointly fused by a Pareto optimization problem
where the mean and the variance of the localization error are
simultaneously minimized. A distinguished feature of the method is
that it assumes a very simple dynamical model of the mobility and
therefore it is applicable to a large number of scenarios providing
good performance.  The main challenge is the characterization of the
statistical information needed to model the Fisher information matrix
and the Pareto optimization problem. The proposed analysis is
validated by Monte Carlo simulations, and the performance is compared
to several Kalman-based filters, commonly employed for localization
and sensor fusion. Simulation results show that the proposed estimator
outperforms the traditional approaches that are based on the extended
Kalman filter when no assumption on the model of motion is used. In
such a scenario, better performance is achieved by the proposed
method, but at the price of an increased computational complexity.
\end{abstract}
%

%
\IEEEpeerreviewmaketitle

\section{Introduction}

In many sensor network applications it is desirable to accurately
estimate the position of mobile wireless nodes~\cite{Patwari2005}.
Relevant applications are vehicular traffic monitoring, asset
tracking, process monitoring, and control of autonomous agents.  As an
example, accurate position information is crucial for emergency
personnel and first responders, see, e.g.,~\cite{Rantakokko2011}. The
commonly used Global Navigation Satellite Systems (GNSSs) provide
position information with an accuracy of about $3$ to $10\:$m in
outdoor scenarios. However, in indoor and electromagnetically-challenged environments, 
such as urban canyons or forests,
the coverage provided by GNSSs is
not sufficient. Therefore, for such environments, several alternative
positioning technologies have been developed, particularly for
wireless indoor localization; for an extensive review see~\cite{Liu2007} and references therein.

The topic of localization using wireless signals has received much
attention in many different research areas. Numerous solutions and
applications have been proposed, including
distributed algorithms for wireless sensor networks~\cite{Patwari2003},
the robust distributed method proposed in~\cite{Khan2010}, and
the technique for passive device-free sensor localization proposed
in~\cite{Xi2011}. The key aspect of cooperation between the nodes of
a wireless network for localization purposes has been investigated 
in \cite{Wymeersch2009}, and experimentally evaluated in 
\cite{ContiEtAl2012}.
The fundamental importance of position information for nodes in wireless sensor networks is highlighted in  \cite{Bertrand&Moonen2013}, where topology-aware estimation algorithms are developed, using techniques from the spectral graph theory field. In this context, distributed gossip algorithms for sensor networks localization have been studied in \cite{RabbatEtAl2005} and \cite{KhanEtAl2009}. In particular, \cite{RabbatEtAl2005} shows that kernel averaging techniques for localizing an isotropic source based on measurements by distributed sensors may provide improved robustness and accuracy than least squares estimators. Moreover, in \cite{KhanEtAl2009}, a distributed algorithm is proposed and characterized, which estimates the positions of nodes given a minimal number of anchors and using only data-exchange between neighboring nodes.

The robustness with respect to range-measurement outliers due to non-line of sight conditions has been analyzed in \cite{HammesEtAl2009}, which introduces a robust extended Kalman filter for locating a mobile terminal node in wireless networks. Moreover, in \cite{WangEtAl_TIE_2012}, a Bayesian algorithm is proposed for self-localization and tracking of a mobile node through ranging with known-position anchors. The algorithm is robust to NLOS propagation by combining a Markov model for sight-state estimation and a particle filter for location estimation, with a simple general motion model. Experiments using the IEEE 802.15.4a chirp-spread-spectrum ranging technology show accuracy of approximately 2 m in an indoor environment with varying NLOS and LOS conditions. In \cite{ZachariahEtAl2014}, the concept of schedule-based network localization is introduced, which enables self-localization of mobile nodes, as well as localization of the entire network, without communication overhead.

%

In the context of wireless short range indoor
positioning, Pulse-based Ultra-Wideband (UWB) techniques are of
particular interest due to numerous good qualities, such as a fine
time resolution (which allows for a ranging accuracy of the order of
centimeters when used in conjunction with time-of-arrival methods) a
resilience to multipath propagation effects of the wireless channel,
and a low-power device implementation~\cite{Gezici2005}. Recent
results on the fundamental limits for wideband radio localization of
static nodes have been presented in~\cite{Win2010}. For the case of
mobile cellular systems, a performance analysis has been presented
in~\cite{Spirito2001}.  The fundamental limitations of mobile
localization have been extensively studied in~\cite{Ristic2003} for
bearings-only measurements in the target tracking field, and
in~\cite{Bergman} in the terrain-aided navigation field.  In the
context of sensor networks node localization, the Cram\'er-Rao lower
bound (CRLB) has been evaluated for the case of
received-signal-strength (RSS) range measurements
in~\cite{Arienzo2009} and in~\cite{Cheung2003}. Therein, the RSS model
has been linearized with only ranging measurements and without using
any velocity and orientation information.

\IEEEpubidadjcol

To improve the performance of a positioning system in terms of
reliability of the estimates (integrity), accuracy, and availability,
it is appealing to process information obtained from a number of
sensors by means of fusion
techniques~\cite{SperanzonFischione+08,Gustafsson2010}. These
techniques involve the processing of different information sources to
overcome the fundamental limitations of sensor measurements such as
GNSSs, inertial navigation systems (INSs), odometry, and local radio
technologies. An extensive survey of the most common information
sources and sensor fusion approaches is provided in~\cite{Skog2009} in
the context of automotive positioning.  Concerning information fusion
for UWB, the commonly employed techniques are based on the extended
Kalman filter (EKF),~\cite{Kailath}, or other non-linear filtering
approaches such as particle filters methods~\cite{Jourdan2005}.
In~\cite{Hol2009}, a 6 degrees-of-freedom tracking system is
presented, which performs sensor fusion on a UWB distance-measuring
device and an inertial sensor consisting of triaxial accelerometers
and gyroscopes. 
In~\cite{DeAngelis_MMS2010}, an UWB sensor fusion
technique based on the complementary filter approach is presented,
where the error states are estimated by an EKF using the UWB ranging
measurements. Subsequently, the estimated errors are fed back to
correct for the inertial navigation system biases, which would
otherwise grow unbounded.

In this paper, 
we investigate a novel sensor fusion technique, based on Pareto optimization,
for node self-localization using heterogeneous information sources. It
is assumed that a node wishing to estimate its own position has
available ranging, speed and orientation information.  We stress here
that the problem that we consider in this paper is self-localization,
which is fundamentally different from the target tracking problem that
is widely studied in the signal processing literature, see
e.g. \cite{Niu_Varshney2012}.  Furthermore, the fundamental
performance limitation of the linear fusion is investigated in this
paper. This investigation is a substantial extension of our earlier work in \cite{DeAngelisIFAC} and \cite{DeAngelisDCOSS}, where the performance of the Pareto optimization were not fully investigated and no derivation of the CRLB were considered. The
\emph{posterior} CRLB~\cite{Tichavsky1998} and the \emph{parametric}
CRLB~\cite{Bergman} are characterized for both any generic trajectory
of the mobile node and for a specific trajectory. 
Since ranging information gives an
error with low bias and a high variance, and measurements of the speed
and absolute orientation (dead-reckoning) give an error with a low
variance and a high bias, the overall information is jointly processed
to overcome the individual limitations of ranging, speed and
orientation measurements. By numerical simulations and by evaluating
the CRLB, we show that our new method may outperform existing
solutions, including several commonly employed techniques based on the
Kalman filter. However, the price payed is a higher computational
complexity.

This paper has several distinguished features compared to the related work
mentioned above. Specifically, our system model is original because
the mobile node position estimation is performed by ranging
measurements, with respect to fixed position nodes, together with
velocity and orientation measurements without any linearization or simplification. 
For the derivation of the
estimator, we assume a very simple dynamical model of the movement of
the mobile node. This model allows for flexibility, robustness,
low-complexity implementation and works well in several different
scenarios, as we show later. Only local processing at such a node is
used,
thus enabling a completely distributed strategy. 
Furthermore, the estimation is performed in a cooperative fashion, 
in the sense that operation of the estimator at the mobile node relies on
the cooperation of fixed-position nodes acting as responders during ranging measurements.

Therefore, our method is of a general
nature and can be applied to any motion scenario. Compared to,
e.g.,~\cite{Arienzo2009} and~\cite{Cheung2003}, we investigate the
CRLB without any linearization of the ranging model and we consider
simultaneously ranging, velocity and orientation measurements, which
were not considered therein. 
Moreover, in this paper, we address 
one of the main challenges for developing sensor fusion methods, 
namely the statistical characterization of the estimation error.
Based on this characterization, a sensor fusion method is derived by solving a Pareto optimization problem, where a tradeoff between the variance of the estimation error and its bias is exploited. The Pareto optimization approach was first proposed in our initial study in this context in \cite{DeAngelisDCOSS}.

The remainder of this paper is organized as follows:
The problem is formulated in Section~\ref{problem_form}.  Then, the proposed
sensor fusion method is derived in Section~\ref{derivation}. Furthermore, the
fundamental limits of an estimator that fuses ranging, velocity and
orientation measurements are investigated in
Section~\ref{section:CRLB} by means of the CRLB.  Numerical simulation
results are presented in Section~\ref{simulations}. Finally,
conclusions are drawn in Section~\ref{conclusion}.

\subsection{Notation} \label{notation} We denote real
$n$-dimensional vectors with lowercase boldface letters, such as $\x$,
and $\R^{n \times m}$ matrices with uppercase boldface letters, such
as $\A$. The superscript $(\cdot)^{T}$ indicates the transpose of a
matrix. Given two matrices $\A$ and $\B$, the inequality $\A \preceq
\B$ denotes that the matrix $\B - \A$ is positive semidefinite. Given
a scalar function $f(\x):\R^n \rightarrow \R$, $ \nabla_{\x}
f(\x)=\left[\frac{d f(\x)}{dx_1}, \ldots, \frac{d
    f(\x)}{dx_n}\right]^T$. Given a vector function $\f(\x) :\R^n
\rightarrow \R^n $, we use the gradient matrix definition $\nabla_{\x}
\f(\x) =\begin{bmatrix} \nabla f_1(\x) & \ldots & \nabla
f_n(\x) \end{bmatrix}, $ which is the transpose of the Jacobian
matrix.
We indicate with $\tilde{\x}$ a noisy measurement of $\x$ and with
$\hat{\x}$ an estimator of $\x$. The statistical expectation operator
is denoted by $\E\left\{ \cdot \right\}$. $p(\x)$ denotes a
probability density function (pdf) of the random vector $\x$.

\section{Problem formulation}
 \label{problem_form}

Consider a system where a mobile node, which we denote as
\emph{master}, needs to estimate its own planar position $\x_k \in
\R^2$, with $\x_k = \left[ x_{1,k} \quad x_{2,k} \right]^T$, where $k$
is the discrete time index. To do so, the master measures its distance
with respect to $M$ devices, which we denote as \emph{slaves}, with $M
\geq4 $. We do not assume any a-priori information on the mobility of
the master. It is allowed to move by following a linear trajectory
with constant velocity as well as a random trajectory with random
acceleration. An example of such a system is provided by the in-house
developed experimental platform that has been characterized in
\cite{DeAngelis_MMS2010,DeAngelis_UWB_system_journal}, where the
distance measurement is obtained by means of the round-trip-time of
UWB signals. 
This distance measurement approach, also known in the literature 
as two-way time-of-arrival, avoids the need for synchronization
between mobile node and slaves, therefore allowing for asynchronous operation 
and reduced infrastructure requirements. Alternative approaches, such as 
one-way time-of-arrival and time-difference-of-arrival, suffer from clock 
skew and require additional infrastructure for precise clock 
synchronization \cite{GholamiEtAl2013}. 

\begin{figure}[!tb]
\centering
\resizebox{0.90\linewidth}{!}{\includegraphics{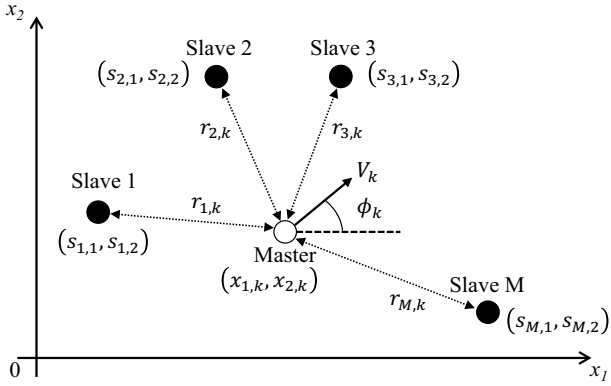}}
  \caption{Model of the localization system. The slave nodes are
    placed at fixed and known positions. At time $k$, the
    unknown-position master node measures its distance $r_{i,k}$ with
    respect to each slave $i$ by measuring the round-trip-time of UWB
    pulses. Speed $V_k$ and orientation $\phi_k$ of the master are
    measured using on-board sensors.}
  \label{F:system_model}
\end{figure}

We assume that the master is equipped with UWB
transmission and sensors that measure its speed and absolute
orientation. 
As an example, speed sensors might be implemented by means of odometry in vehicular or mobile-robotics applications, such as those modeled in~\cite{Mourikis2006}. Furthermore, information about absolute orientation could be obtained from a heading sensor, for example a  compass or magnetometer, as in \cite{Georgiev2004}.
The diagram in Fig.~\ref{F:system_model} shows a representation of the system
model.
Furthermore, we assume that the slaves are located in fixed and known positions
that are denoted as $s_{i,j}$, with $i=1,\ldots, M$ and $j=1,2$.  We
denote the ranges, namely the distances between the master and each
slave at time instant $k$, by $r_{i,k}$, $i=1,\ldots, M$. The
measurements of the ranges, denoted as $\tilde{r}_{i,k}$, are affected
by additive zero-mean Gaussian noise $w_{r,i,k}$:
\begin{align}
\tilde{r}_{i,k} = r_{i,k} + w_{r,i,k}\,. \label{e:range_expression}
\end{align}
The variance of $w_{r,i,k}$ depends on the range, according to the
following exponential model that has been derived from real-world
round-trip-time UWB measurements in~\cite{DeAngelis_MMS2010}:
\begin{align}
\sigma^2_{r,i,k} = \sigma^2_0 \exp(\kappa_{\sigma} r_{i,k})\,,
\label{e:range_variance_model}
\end{align}
where $\sigma^2_0$ and $\kappa_{\sigma}$ are known constants.

Further, we denote the speed and orientation of the master by $V_k$
and $\phi_k$, respectively, whereas $\tilde{V}_k$ is the measurement
of the speed and $\tilde{\phi}_k$ the measurement of the
orientation. These measurements are expressed as
$$ \tilde{V}_k = V_k + w_{V,k} \,, \quad \tilde{\phi}_k = \phi_k +
w_{\phi,k}
$$ where the noise terms $w_{V,k}$ and $w_{\phi,k}$ are zero-mean
Gaussian random variables with known variances $\sigma_{V}^2$ and
$\sigma_{\phi}^2$, respectively. 


In this paper, assuming the measurement models described above, we propose a novel method to combine and improve existing estimators which are well known and commonly used.
Specifically, we use a so called loosely-coupled approach, in which we fuse estimates already available from ranging measurements, with the Weighted Least Squares (WLS) method, and from velocity and orientation measurements, by performing dead-reckoning.
Note that, in order to achieve the best possible accuracy, a tightly-coupled approach is necessary, where information from the ranging, velocity and orientation sensors are fused directly without preprocessing.
However, the loosely-coupled approach is motivated by the reduced computational complexity as well as by the increased robustness and modularity that it offers. 

This paper has two goals: First, to propose an highly accurate estimator of the position of the master by processing the results provided by existing ranging-only WLS and dead reckoning estimators. Second, to understand the fundamental limitations when estimating the position of the master by the fusion of information from these estimators.  In the remainder of this section, we start by formulating the estimation problem that leads to the derivation of the proposed estimator.

The estimate
of the master's position at time $k+1$ by using measurements available
at that time is denoted by $ \hat{\x}_{k+1 \mid k+1} =
\left[\hat{x}_{1, k+1 \mid k+1} \quad \hat{x}_{2, k+1 \mid
k+1}\right]^T$. We propose that it is derived as follows:
\begin{align}
\hat{x}_{1, k+1 \mid k+1} & = \alpha_{1,k}
\hat{x}_{1,k+1}^{(r)} + \beta_{1,k} \hat{x}_{1,k+1}^{(v)}\,,
    \label{e:model_old_alpha}\\
\hat{x}_{2, k+1 \mid k+1} & = \alpha_{2,k}
\hat{x}_{2,k+1}^{(r)} + \beta_{2,k}
\hat{x}_{2,k+1}^{(v)}\,,\label{e:model_old_alpha_y}
\end{align}
where $\hat{\x}_{k+1}^{(r)}=\left[\hat{x}_{1,k+1}^{(r)} \quad
  \hat{x}_{2,k+1}^{(r)}\right]^T$ is the position estimate based only
on the ranging measurements, and
$\hat{\x}_{k+1}^{(v)}=\left[\hat{x}_{1,k+1}^{(v)} \quad
  \hat{x}_{2,k+1}^{(v)}\right]^T$ is the position estimate based on
the dead reckoning block, which processes the position estimate at the
previous time step and the on-board speed and orientation sensors. The
terms $\alpha_{1,k}$, $\alpha_{2,k}$, $\beta_{1,k}$ and $\beta_{2,k}$ are the sensor fusion design
parameters that need to be optimally chosen.
Our estimation problem is separable on the $x_1$ and $x_2$ axes,
in the sense that the estimates on the $x_2$ axis does not affect the
$x_1$ axis, and vice versa. Therefore in the following we will provide
derivations only for the $x_1$ component, because the derivations for
the $x_2$ component are similar to those for the $x_1$ component.

Typically, the dead reckoning block provides us with an estimate
having the following expression:
\begin{align}
 \hat{x}_{1,k+1}^{(v)} & = \hat{x}_{1,k \mid k} + \tilde{V}_k T
 \cos{\tilde{\phi}_k} \label{e:est_x} \,.
\end{align}
Note that we do not make the assumption of a linear motion model for
the dead-reckoning block.  Furthermore, the dead-reckoning estimate is
biased because the orientation measurement appears as the argument of
a cosinus. It follows that the estimate~\eqref{e:model_old_alpha} is biased.
This motivates us to model our estimator~\eqref{e:model_old_alpha} as
\begin{align}
\hat{x}_{1,k+1 \mid k+1} \triangleq x_{1,k+1} +
w_{x_{1,k+1}}\,, \label{e:general_expr}
\end{align}
where $w_{x_{1,k+1}}$ is the error in the position estimate. This
error has a non zero average, since the dead reckoning block gives a
biased estimate.

It is possible to write a simplified form of Equations \eqref{e:model_old_alpha}, by first noting that
\begin{align*}
\E \left\{ \hat{x}_{1, k+1 \mid k+1} \right\} & =
\alpha_{1,k} x_{1,k+1} + \beta_{1,k} x_{1,k+1} \\
& \quad + \alpha_{1,k} \E \left\{ w_{r,k+1} \right\} \\
& \quad + \beta_{1,k} \E \left\{ w_{v,k+1} \right\}
+ \beta_{1,k} \E \left\{ w_{x,k} \right\} \\
& = x_{1,k+1} + \E \left\{ w_{x,k+1} \right\}
\end{align*}
where $w_{r,k+1}$ and $w_{v,k+1}$ are the errors in the estimate obtained from the ranging block and from the dead-reckoning block, respectively. Now, in order to have $\alpha_{1,k} x_{1,k+1} + \beta_{1,k} x_{1,k+1} = x_{1,k+1}$, it must be $\alpha_{1,k} + \beta_{1,k} = 1$ and therefore 
\begin{align}
\alpha_{1,k} = 1 - \beta_{1,k} \,. \label{e:alpha_beta_relation}
\end{align}
Hence, by substituting \eqref{e:alpha_beta_relation} in \eqref{e:model_old_alpha}, we obtain
\begin{align}
\hat{x}_{1, k+1 \mid k+1} & = \left(1-\beta_{1,k}\right)
\hat{x}_{1,k+1}^{(r)} + \beta_{1,k} \hat{x}_{1,k+1}^{(v)}\,,
    \label{e:model_old}
\end{align}
In the following, we will use the simplified expression \eqref{e:model_old} instead of \eqref{e:model_old_alpha}.

From~\eqref{e:general_expr} it follows that the error
is
\begin{align}
w_{x_{1,k+1}} & = \hat{x}_{1,k+1 \mid k+1} -
x_{1,k+1} \label{e:general_expr_error} \\ 
& = \left( 1 - \beta_{1,k}
\right) w_{x_{1,k + 1}}^{(r)} + \beta_{1,k} w_{x_{1,k}}  \notag \\
& \quad +\beta_{1,k} T
\tilde{V}_k \cos\tilde{\phi}_k - \beta_{1,k} T V_k
\cos\phi_k\,, \label{e:estimation_error}
\end{align}
where $w_{x_{1,k + 1}}^{(r)}$ is the error of the estimates obtained
from the ranging.
To develop an estimator for the master's position, we define a cost
function that takes into account the estimator variance and bias
simultaneously. We define the bias term of the estimation error as $
\mu_{w1,k+1} \triangleq \E \left\{ w_{x_{1,k+1}} \right\} \,, $ and
the variance term of the estimation error as $ \sigma^2_{w1,k+1}
\triangleq \E \left\{ \left( \hat{x}_{1,k+1 \mid k+1} - \E \left\{
\hat{x}_{1,k+1 \mid k+1} \right\} \right)^2 \right\} = \E \left\{
\left( w_{x_{1,k+1}} - \E \left\{ w_{x_{1,k+1}} \right\} \right)^2
\right\} \,.  $

We are now ready to formulate a Pareto optimization problem to select
the fusion coefficients at each time $k$:
\begin{align} \label{e:problem2_rho_simple}
\min_{\beta_{1,k}} & \quad \rho_{1,k} \mu_{w1,k}^2 + \left( 1 -
\rho_{1,k}\right) \sigma^2_{w1,k} \\ {\rm s.t.} & \quad \beta_{1,k}
\in \mathscr{B}=[-1,1] \,, \nonumber  \quad \rho_{1,k} \in
\mathscr{R}=[0,1]\,. \nonumber
\end{align}
This problem is motivated by that we would like to reduce as much as
possible both the average of the estimation error, namely the bias due to the dead reckoning block, and the variance of the estimation error. 
The Pareto weighting factor (or scalarization coefficient)
$\rho_{1,k}$ must be chosen for each time $k$ so to trade off the low
bias and high variance of the error of the ranging estimate with the
high bias and low variance of the error of the dead-reckoning
estimate~\cite[Section 4.7.5]{Boyd2004}. Based on the statistical properties of the bias, which we
characterize below, the bias itself may grow unstable with time, which
motivates the constraint $\beta_{1,k} \in \mathscr{B}$. The intuition
is that the minimization of both the average and the variance of the
estimation error at every time $k$ does not ensure that the bias is
stable or decreases over time. Notice that in the special case of
$\rho_{1,k} = 0 \, \forall k$, the cost function is the variance of
the estimation error, and when $\rho_{1,k} = 0.5 \, \forall k$ the
cost function is the mean square error (MSE).

The challenge for such an optimization is the analytical
characterization of the cost function~\eqref{e:problem2_rho_simple},
which we study in the following section.

\section{Derivation of a Sensor Fusion Estimator}
\label{derivation}

\begin{figure}[!tb]
\centering \resizebox{0.99\linewidth}{!}{\includegraphics[trim=10 80
    10 50]{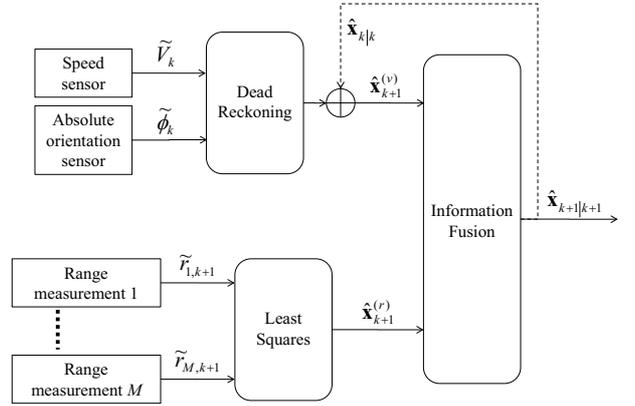}}
  \caption{Sensor fusion system architecture used at a master node for
    estimating its position at time $k+1$ given available information.
    The dead reckoning block gives speed and absolute orientation
    information, whereas the least square block gives ranging
    information as computed with respect to slave nodes. The
    information provided by these two blocks is then fused to provide
    an accurate estimation $\hat{\x}_{k+1|k+1}$.}
  \label{F:fusion_system}
\end{figure}

The architecture of the proposed estimator is shown in
Fig.~\ref{F:fusion_system}.
In the following subsections, we provide a statistical
characterization of the ranging estimate $\hat{\x}_{k+1}^{(r)}$ and
the dead-reckoning estimate $\hat{\x}_{k+1}^{(v)}$.  Then, in
subsection~\ref{solution-po} we solve the optimization problem, thus
achieving the optimal weighting coefficients $\beta_{1,k}$ and
$\beta_{2,k}$ to use in Eq.~\eqref{e:model_old}.
The original contribution of this section is the statistical characterization of the estimation errors, and the optimal weighting derivation in Proposition 3.7. 

\subsection{Ranging estimate statistical characterization} \label{ranging}

In this subsection we give a new statistical characterization of the
estimation error of the ranging estimate. We consider the problem of
finding the first and second order moments of the ranging estimates of
the $x_1$ and $x_2$ components of the position. This requires first a
derivation of the ranging estimates. In the following, for the sake of
simple notation, we drop the $k+1$ indices.


Since it must be that $M \geq 4$ for linear-based ranging estimates,
the master's position estimate $\mathbf{\hat{x}}^{(r)}$ may be
obtained by following a commonly employed strategy (see,
e.g.,~\cite{Guvenc2010}) as the solution of $\A
\hat{\x}^{(r)}=\tilde{\s}$, where $\A$ and $\tilde{\s}$ are easily
seen to be
\begin{align}
\A = \left[ \begin{array}{cc} 2\left(s_{1,1} - s_{M,1}\right) &
    2\left(s_{1,2} - s_{M,2}\right) \\ \vdots & \vdots
    \\ 2\left(s_{M-1,1} - s_{M,1}\right) & 2\left(s_{M-1,2} -
    s_{M,2}\right)
                     \end{array}
        \right] \,,
\label{e:H}
\end{align}
\begin{align}
\mathbf{\tilde{s}} = & \left[ \begin{array}{c} r_M^2 - r_1^2, \ldots,
    r_M^2 - r_{M-1}^2
             \end{array}
\right]^T + \mathbf{a} + \mathbf{w_s} \,,
\end{align}
where the constant vector $\a$ is given by the following function of
the known coordinates of the slaves
\begin{align*}
\a = \left[ \begin{array}{c} s_{1,1}^2 - s_{M,1}^2 + s_{1,2}^2 -
    s_{M,2}^2 \\
    \ldots \\
    s_{M-1,1}^2 - s_{M,1}^2 + s_{M-1,2}^2 -
    s_{M,2}^2
    \end{array}
\right] \,,
\end{align*}
and $\mathbf{w_s}$ is a random vector given by
\begin{align} \label{e:WLS_error_w_s}
\mathbf{w_s} \triangleq \left[ 
  \begin{array}{c} w_{r_M}^2 + 2r_M w_{r_M} - w_{r_1}^2 - 2r_1 w_{r_1} \\
    \ldots \\
    w_{r_M}^2 + 2r_M w_{r_M} - w_{r_{M-1}}^2 - 2r_{M-1} w_{r_{M-1}}
  \end{array}
\right] \,. 
\end{align}

We derive the ranging estimate by the weighted least squares (WLS)
technique, since ranging measurements are affected by unequal
statistical errors across the various links. Then, the solution to the
following classical WLS optimization problem gives the ranging
estimates: $ \min_{\hat{\x}^{(r)}} \left\| \W^{\frac{1}{2}} \left(
\mathbf{\tilde{s}} - \A \hat{\x}^{(r)} \right) \right\|$, where $\W$
is a positive-definite weighting matrix introduced for the purpose of
emphasizing the contribution of those range measurements that are
deemed to be more reliable~\cite{Kay}, and $\| \cdot \|$ is the
Euclidean norm. A common choice for $\W$ is $\mathbf{R}^{-1}$, namely
the inverse of the noise covariance matrix:
\begin{align}
\mathbf{R} \triangleq \E \left\{ \left( \w_s - \E \left\{ \w_s
\right\} \right) \left( \w_s - \E \left\{ \w_s \right\} \right)^T
\right\}\,.
\label{e:cov_definition}
\end{align}
Such a choice is motivated by that it would give a maximum likelihood estimator in the case of Gaussian errors.
Thus the well-known solution to the WLS problem is
\begin{align}
\hat{\x}^{(r)} = \left( \A^T \W \A\right)^{-1} \A^T \W \tilde{\s} \,,
\label{e:positioning_estimate}
\end{align}
where $\W=\mathbf{R}^{-1}$. 

We are now ready to introduce the new part of this subsection.
From~\eqref{e:positioning_estimate} we can
characterize the first and second order moments of the error in the
ranging estimate. First, however, we need the expression of the
inverse of the error covariance matrix. This inverse matrix has a simple expression and is provided by the following lemma:

\begin{lemma} \label{lemma:cov-mat}
Consider the matrix $\mathbf{R}$ as defined in
\eqref{e:cov_definition}. The matrix has full rank, and the inverse is
\begin{align*}
\mathbf{R}^{-1} = \left( \I - \frac{1}{1+q} \G \1 \1^T \right) \D^{-1}
\,,
\end{align*}
where $\I$ is the $(M-1)\times(M-1)$ identity matrix, $\D$ is a
$(M-1)\times(M-1)$ diagonal matrix whose entries are $\D_{ll} = 4
r_l^2 \sigma_{w_{r_l}}^2 + 2\sigma_{w_{r_l}}^4$, $l=1,\ldots, M-1$,
$\G=p\D^{-1}$ is a $(M-1)\times(M-1)$ matrix with $p = 4 r_M^2
\sigma_{w_{r_M}}^2 + 2\sigma_{w_{r_M}}^4$, $q=\sum\limits_{i=1}^{M-1}
p/\D_{ii}$, and $\1$ is the all ones vector.
\end{lemma}
\begin{IEEEproof}
The lemma follows from tedious algebraic computations, the matrix
inversion lemma, and the Woodbury identity~\cite{Kay}.
\end{IEEEproof}

Based on Eq.~\eqref{e:positioning_estimate} and the previous lemma, we
are now in the position of deriving the expectation and the variance
of the ranging estimation error $\w_r$, defined as $ \w_r \triangleq
\hat{\x}^{(r)} - \x = \left[ w_{x_{1}}^{(r)} \quad
  w_{x_{2}}^{(r)} \right]^T\,.  $

\begin{lemma} \label{Proposition_mean_ranging}
The expectation of the ranging estimation error is given by:
\begin{align*}
\E\left\{ \w_r \right\} = & \left(\A^T\W\A\right)^{-1}\A^T \W \\
& \times \left( \1 \sigma^2_{w_{r_M}} - \left[ \sigma^2_{w_{r_1}}, \hdots,
  \sigma^2_{w_{r_{M-1}}} \right] ^T \right).
\end{align*}
\end{lemma}
\begin{IEEEproof}
See Appendix \ref{appendix_mean_ranging}.
\end{IEEEproof}

Furthermore, we have the following result:
\begin{lemma} \label{Proposition_correlation_ranging}
The correlation of the range estimation error is given by
\begin{align} \label{e:somr}
\E\left\{\w_r^2\right\}
= {\left(\A^T\W\A\right)}^{-1}\A^T \W\C
\W^T\A{\left(\A^T\W\A\right)}^{-1}\,,
\end{align}
where $\C$ is a matrix whose diagonal elements are $ \C_{ll}
= 3\sigma^4_{w_{r_M}} + 4 r_M^2 \sigma^2_{w_{r_M}} + 3
\sigma^4_{w_{r_l}} + 4 r_l^2 \sigma^2_{w_{r_l}} -
2\sigma^2_{w_{r_M}}\sigma^2_{w_{r_l}} \,, $ with $l=1, \dots , M$, and
the off-diagonal elements are $ \C_{lj}
= 3\sigma^4_{w_{r_M}} - \sigma^2_{w_{r_M}} \sigma^2_{w_{r_j}} + 4
r_M^2 \sigma^2_{w_{r_M}} - \sigma^2_{w_{r_l}}\sigma^2_{w_{r_M}} +
\sigma^2_{w_{r_l}}\sigma^2_{w_{r_j}}\,, $ with $l \neq j$.
\end{lemma}
\begin{IEEEproof}
See Appendix \ref{appendix_correlation_ranging}.
\end{IEEEproof}

This concludes the efforts to characterize the statistics of the
ranging estimation error. In the following subsection, we focus on the
statistics of the dead-reckoning estimation.

\subsection{Dead-reckoning estimate statistical characterization}

In this subsection, we give a new characterization of the estimation errors of the dead-reckoning block. 
We have the following results for the first and second order moments
of the dead-reckoning estimate:
\begin{lemma} \label{Proposition_1}
Let $\tilde{\phi}(k)$ be Gaussian, and assume that $\tilde{V}( k)$ and
$\tilde{\phi} ( k)$ are statistically independent. Then
\begin{align}
\E \left\{ \tilde{V}_k \cos\left(\tilde{\phi}_k\right) \right\} & =
V_k \cos\left(\phi_k\right) e^{-\frac{\sigma_{\phi}^2}{2}}
\,, \label{e:expectation_inertial} \\ \E
\left\{\tilde{V}^2_k\cos^2\left(\tilde{\phi}_k\right)\right\} & =
\sigma^2_V \left( \frac{1}{2} + \frac{1}{2}
\cos\left(2\phi_k\right)e^{-2\sigma^2_\phi} \right)\,.
\label{e:2nd_order_inertial}
\end{align}
\end{lemma}
\begin{IEEEproof}
See Appendix \ref{appendix_Proposition_1}.
\end{IEEEproof}
In the following subsection, we use these results to characterize the
estimation bias.

\subsection{Estimation Bias}

We can now put together the statistical characterization of the
ranging and dead-reckoning estimations. We have the following results
that give the terms $\mu_{w1,k}$ and $\sigma^2_{w1,k}$ in
Eq.~\eqref{e:problem2_rho_simple}:
\begin{lemma} \label{Proposition_3}
Consider the estimation error~\eqref{e:general_expr_error}. The bias
is
\begin{align}
\lefteqn{\mu_{w1,k+1}=  \E \{w_{x_{1,k+1}}\} = \left( 1 - \beta_{1,k} \right)\E \left\{ w_{x_{1,k + 1}}^{(r)} \right\}} & \nonumber \\
& + \beta_{1,k} \E \{w_{x_{1,k}}\} + \beta_{1,k} T V_k \cos\left(\phi_k\right) \left(e^{-\frac{\sigma_{\phi}^2}{2}} - 1 \right) \,.  \label{e:mean_noise}
\end{align}
\end{lemma}
\begin{IEEEproof}
See Appendix \ref{appendix_Proposition_3}.
\end{IEEEproof}

\begin{remark}
From the previous lemma, we see that the average of the estimation
error at time $k+1$ depends on the bias of time $k$. To avoid an
accumulation of the bias, we need to impose a condition on
$\beta_{1,k}$. Thus, the absolute value of the average of the
estimation error must be non-expansive, which can be easily achieved
when $|\beta_{1,k}| \in [0,1]$, and thus we have $\mathscr{B}=[-1,1]$
in the Pareto optimization problem \eqref{e:problem2_rho_simple}.
\end{remark}

\begin{lemma} \label{Proposition_4}
Consider the estimation error~\eqref{e:general_expr_error}. Then, the
second order moment is
\begin{align} \label{e:variance_term}
\E\left\{w_{x_{1,k+1}}^2 \right\} = \beta_{1,k}^2 a_k + 2 \beta_{1,k}
b_k + \E \left\{\left(w_{x_{1,k+1}}^{(r)}\right)^2\right\}
\end{align}
where
\begin{align}
\lefteqn{
a_k = \E\left\{\left(w_{x_{1,k+1}}^{(r)}\right)^2\right\} +
\E\left\{w_{x_{1,k}}^2\right\} + T^2 \E
\{\tilde{V}^2_k\cos^2(\tilde{\phi}_k)\} 
} & \nonumber \\
& + T^2 V^2_k\cos^2(\phi_k)
\left(1 - 2 e^{-\frac{\sigma_{\phi}^2}{2}} \right) 
- 2 \E \{w_{x_{1,k}}\} \E \{w_{x_{1,k + 1}}^{(r)}\} \nonumber \\ 
&  - 2 \E \{w_{x_{1,k + 1}}^{(r)}\} T V_k\cos\left(\phi_k\right) \left(
e^{-\frac{\sigma_{\phi}^2}{2}} - 1 \right) \nonumber \\ 
& + 2 \E \{w_{x_{1,k}}\} T V_k\cos\left(\phi_k\right) \left(
e^{-\frac{\sigma_{\phi}^2}{2}} - 1 \right)\,, \label{e:a_k} \\ 
\lefteqn{b_k = - \E\left\{\left(w_{x_{1,k+1}}^{(r)}\right)^2\right\} 
+ \E \{w_{x_{1,k}}\} \E \{w_{x_{1,k + 1}}^{(r)}\}} & \nonumber \\
& + \E \{w_{x_{1,k + 1}}^{(r)}\} T V_k\cos\left(\phi_k\right) \left(
e^{-\frac{\sigma_{\phi}^2}{2}} - 1 \right)\,. \label{e:b_k}
\end{align}
\end{lemma}
\begin{IEEEproof}
The result follows from Lemma~\ref{Proposition_mean_ranging},
Eq.~\eqref{e:somr}, and Proposition~\ref{Proposition_1}.
\end{IEEEproof}

Finally, to derive an expression for the variance of the estimation
error $\sigma^2_{w1,k}$ that we need in~\eqref{e:problem2_rho_simple},
we define the variables $v_r$, $v_{x_1}$ and $v_v$ as
\begin{align*}
v_r & \triangleq w_{x_{1,k+1}}^{(r)} - \E \left\{w_{x_{1,k+1}}^{(r)}
\right\}, \, v_{x_1} \triangleq w_{x_{1,k}} - \E \left\{ w_{x_{1,k}}
\right\}, \, \\
v_v & \triangleq T \tilde{V}_{k}
\cos\left(\tilde{\phi}_{k}\right) - T V_{k} \cos \left(\phi_{k}\right)
e^{-\frac{\sigma_{\phi}^2}{2}} \,.
\end{align*}
Notice that $v_r$, $v_{x_1}$ and $v_v$ are independent; we denote
their variances as $\sigma_{v_r}^2$, $\sigma_{v_{x_1}}^2$ and
$\sigma_{v_v}^2$ respectively. Also, note that $\E\left\{v_r\right\} =
\E \left\{v_{x_1}\right\} = 0$.
By substituting the expression for the error bias~\eqref{e:mean_noise}
in Eq.~\eqref{e:variance_term} we obtain
\begin{align}
\sigma^2_{w1,k+1} = & \E \left\{ \left( \left( 1 - \beta_{1,k} \right)
w_{x_{1,k + 1}}^{(r)} + \beta_{1,k} w_{x_{1,k}} 
\right. \right. \notag \\
& \left. \left. + \beta_{1,k} T \tilde{V}_{k} \cos\left(\tilde{\phi}_{k}\right) - \beta_{1,k} T V_{k}
\cos \left(\phi_{k}\right) \right. \right. \nonumber \\ 
& \left. \left. - \left( 1 - \beta_{1,k} \right) \E \left\{ w_{x_{1,k +
    1}}^{(r)} \right\} - \beta_{1,k} \E \left\{ w_{x_{1,k}} \right\} 
\right. \right. \notag \\
& \left. \left. + \beta_{1,k} T V_{k} \cos \left(\phi_{k}\right)
e^{-\frac{\sigma_{\phi}^2}{2}} \right)^2 \right\} \nonumber \\
= & \E \left\{ \left( 1 - \beta_{1,k} \right)^2 v_r^2 + \beta_{1,k}^2
v_{x_1}^2 + \beta_{1,k}^2 v_v^2 
\right. \nonumber \\ 
& \left. 
+ 2 \left( 1 - \beta_{1,k} \right) \beta_{1,k} v_r v_{x_1} + 2 \left( 1 -
\beta_{1,k} \right) \beta_{1,k} v_r v_v 
\right. \nonumber \\ 
& \left. + 2 \beta_{1,k}^2 v_{x_1} v_v
\right\}
\nonumber \\ 
= & \left( 1 - \beta_{1,k} \right)^2 \sigma_r^2 + \beta_{1,k}^2
\sigma_x^2 + \beta_{1,k}^2 \sigma_v^2 \,,
\label{e:variance_noise}
\end{align}
where we have used that $v_r$, $v_{x_1}$ and $v_v$ are independent and
$\E \left\{v_r\right\} = \E \left\{v_{x_1}\right\} = 0$.

\subsection{Pareto Optimization} \label{solution-po}

In this section, we put together the results of the previous sections
to derive a position estimator by solving the optimization problem
in~\eqref{e:problem2_rho_simple}.  Lemmata~\ref{Proposition_3}
and~\ref{Proposition_4} give us the analytical expression of the cost
function of problem~\eqref{e:problem2_rho_simple}. Thus, we have the
following result, which is one of the core contributions of this
paper:
\begin{proposition} \label{Proposition_5}
Consider optimization problem~\eqref{e:problem2_rho_simple}. Let the
mean and the variance of the estimation
error~\eqref{e:general_expr_error} be computed by
Lemmata~\ref{Proposition_3} and~\ref{Proposition_4}. Then, the optimal
solution for a fixed Pareto weighting factor $\rho_{1,k}$ is
\begin{align}\label{e:problem2_solution}
\beta_{1,k}^*(\rho_{1,k})=\max\left(-1,\min \left(\xi,
1\right)\right)\,,
\end{align}
with
$$ \xi = \frac{2\left( 1 - \rho_{1,k}\right)\sigma_{v_{r}}^2 - 2
  \rho_{1,k} \gamma \E\left\{ w_{r,k+1}\right\}} {2\left( 1 -
  \rho_{1,k}\right) \eta + 2 \rho_{1,k} \gamma^2} \,,
$$ where
\begin{align*}
\eta & \triangleq \sigma_{v_{r}}^2 + \sigma_{v_{x}}^2 + \sigma_{v_v}^2
\,, \\
\gamma & \triangleq - \E \left\{ w_{r,k+1} \right\} + \E
\left\{ w_{x,k} \right\} + T V_k \cos\left(\phi_k\right) \left(
e^{-\frac{\sigma_{\phi}^2}{2}} - 1 \right) \,.
\end{align*}
\end{proposition}
\begin{IEEEproof}
See Appendix \ref{appendix_problem2_solution}.
\end{IEEEproof}
As a particular case of the previous proposition is given by the MSE, when the optimal solution
of problem~\eqref{e:problem2_rho_simple} is obtained with the Pareto weighting factor 
$\rho_{1,k}=0.5$ $\forall k$ and results 
\begin{align}
\beta_{1,k}^*=\max\left(-1,\min \left(- \frac{b_k}{a_k},
1\right)\right)\,, \label{e:special_case_sol}
\end{align}
where $a_k$ and $b_k$ are given by Eq.~\eqref{e:a_k} and
Eq.~\eqref{e:b_k}, respectively. In general, the optimal value of
$\beta_{1,k}$ in~\eqref{e:problem2_solution} depends on the scalarization coefficient 
$\rho_{1,k}$ (see~\cite[Section 4.7]{Boyd2004}). The best value of such a coefficient is found by building
the Pareto trade-off curve and selecting the ``knee-point'' on this
curve~\cite{Boyd2004}. Thus, we choose $\rho_{1,k}^*$ such that
$\mu_{w1,k}$ and $\sigma^2_{w1,k}$ computed in $\beta_{1,k}^*
(\rho_{1,k}^*)$ are $\sigma^2_{w1,k} \simeq \mu_{w1,k}^2$. This is
given by the solution to the following further optimization problem
\begin{align} \label{e:Pareto_rho_selection}
\min_{\rho_{1,k}} \left( \sigma^2_{w1,k}\left(\beta_{1,k}^*
\left(\rho_{1,k}\right)\right) - \mu_{w1,k}^2 \left(\beta_{1,k}^*
\left(\rho_{1,k}\right)\right) \right)^2 \,,
\end{align}
where we have evidenced that $\mu_{w1,k}$ and $\sigma^2_{w1,k}$ are
computed in $\beta_{1,k}^* \left(\rho_{1,k}\right)$, according to
Eqs.~\eqref{e:mean_noise} and~\eqref{e:variance_noise}. This problem
is highly non-linear, but simple and standard numerical procedures,
based on a discrimination of $\rho_{1,k}$, can be
employed to compute approximately the optimal Pareto coefficient
$\rho_{1,k}^*$.  
The standard numerical procedure is described in~\cite[Section 4.7.5]{Boyd2004}. Such a procedure consists of first defining a set of values that $\rho_{1,k}$ can assume, then for each such value the procedure does the following steps:
\begin{enumerate}
\item compute $\beta^*(\rho_{1,k})$ via \eqref{e:problem2_solution} for the current value of $\rho_{1,k}$, 
\item use the value of $\beta^*(\rho_{1,k})$ computed in 1) and plug it in~\eqref{e:mean_noise} and in~\eqref{e:variance_noise} to evaluate $\mu_{w1,k}^2 \left(\beta_{1,k}^*
\left(\rho_{1,k}\right)\right)$ and $\sigma^2_{w1,k}\left(\beta_{1,k}^* \left(\rho_{1,k}\right)\right)$;
\item use the values $\mu_{w1,k}^2 \left(\beta_{1,k}^*
\left(\rho_{1,k}\right)\right)$ and $\sigma^2_{w1,k}\left(\beta_{1,k}^* \left(\rho_{1,k}\right)\right)$ computed at step 2) to compute
\begin{align} \label{e:Pareto_rho_selection_specific}
\left(\sigma^2_{w1,k}\left(\beta_{1,k}^* \left(\rho_{1,k}\right)\right) - \mu_{w1,k}^2 \left(\beta_{1,k}^*
\left(\rho_{1,k}\right)\right) \right)^2 \,,
\end{align}
and store such a value. 
\end{enumerate}
The approximate optimum $\rho_{1,k}^*$ is the one which corresponds to the minimum of the stored values of~\eqref{e:Pareto_rho_selection_specific}. Such a $\rho_{1,k}$ corresponds to the "knee point" of the Pareto trade off curve~\cite[Section 4.7.5]{Boyd2004}.

Finally, recall that the values of the optimal
$\beta_{2,k}$ and $\rho_{x_{2,k}}$ yielding the $x_2$ component of the
position estimate are obtained similarly to what done for the $x_1$
component.

\section{Cram\'er-Rao Lower Bound} \label{section:CRLB}
In this section, we are interested in investigating the fundamental performance limitation of our estimator proposed in the previous section. To do so, given any estimation problem, the CRLB is a fundamental tool for
performance analysis, since it quantifies the best achievable
mean-square error performance of estimators~\cite{Kay}.  
We wish to derive such a bound to compare it to the performance of our estimator. There are two fundamental approaches: the parametric CRLB and posterior CRLB. 

The parametric CRLB, that we denote as \emph{Par}CRLB, is the
CRLB evaluated for a given state-space trajectory
\cite{Bergman}. Therefore it is equivalent to the problem of
estimating unknown deterministic states considering a zero process
noise in the dynamical system model; for the proof see, e.g., the
derivations in~\cite{Taylor1979}.  In other words, the state is
treated as a deterministic but unknown parameter for the purpose of
the evaluation of the \emph{Par}CRLB.  Therefore, the \emph{Par}CRLB
is simple to evaluate in a numerical simulation environment where the
true trajectory is known, because the equations required for its
calculation do not contain any statistical expectation operator; see,
e. g., \cite{Ristic_book}.  However, due to its nature, the
\emph{Par}CRLB does not provide information about the fundamental
performance limitations in the general case.
For such a case, we need to investigate the posterior CRLB, which we do in the following.

\subsection{Posterior CRLB}
For recursive estimation problems in the Bayesian framework, where the
state is treated as a random variable, the bound is denoted as the
\emph{posterior} CRLB (PCRLB). A recursive formulation of the PCRLB
has been first introduced in~\cite{Tichavsky1998}. In the following,
we derive such a bound for our localization problem.  The core
contribution of this section is summarized in
Proposition~\ref{prop:ublbpd} below.

Differently from the case of the \emph{Par}CRLB, where one is 
interested in calculating the performance bound for a given
trajectory, here we aim at characterizing the fundamental
performance limits, on the average, for all the trajectories belonging
to a class, which is defined by a dynamical model. Therefore we
introduce the state $\p_k \in \R^4$ to model the bi-dimensional
position of the master, its absolute speed and its orientation, that
is $\p_{k}= \left[ x_{1,k} \quad x_{2,k} \quad V_{k} \quad \phi_{k}
  \right]^T $, and we consider the following non-linear state-space
model:
\begin{align}
\p_{k+1} & = \f\left(\p_{k}\right) +
\v_k \label{e:PCRLB_system}\\ \y_{k} & = \h \left( \p_k \right) +
\e_k \label{e:PCRLB_measurement}\,,
\end{align}
where $\v_k = \left[ v_{1,k} \quad v_{2,k} \quad v_{3,k} \quad
  v_{4,k}\right]^T \in \R^{4} \sim \mathcal{N}\left( 0, \Q \right)$ is
the white Gaussian process noise with $\mathbf{Q}=\textrm{diag}\left(
\sigma_1^2, \, \sigma_2^2, \, \sigma_3^2, \, \sigma_{4}^2 \right) $, where it is natural to assume that $\sigma_1=\sigma_2$, namely that the process noise on the two coordinates has same variance, 
and $\e_{k} \in \R^{M+2} \sim \mathcal{N}\left( 0, \mathbf{R_k}
\right)$ is the white Gaussian measurement noise, which is independent
of $\v_k$, with $\mathbf{R_k}=\textrm{diag}\left(\sigma_{r,1,k}^2,
\ldots, \sigma_{r,M,k}^2, \sigma_V^2, \sigma_{\phi}^2\right)$.  
Simple algebra gives that the state update and measurement 
functions are, respectively,
\begin{align*}
\f\left(\p_{k}\right) = & \left[
\begin{array}[c]{c}
x_{1,k} + T V_k \cos{\phi_k} \\ x_{2,k} + T V_k \sin{\phi_k} \\ V_k
\\ \phi_k
\end{array}
\right] \,, \\
\h\left( \p_k \right)= & \left[
\begin{array}[c]{c}
h_1 , \, \ldots , \, h_M , \, h_{M+1} , \, h_{M+2}
\end{array}
\right]^T  \\
= & \left[
\begin{array}[c]{c}
\sqrt{(s_{1,1}-x_{1,k})^2+(s_{1,2}-x_{2,k})^2} \\ \vdots
\\ \sqrt{(s_{M,1}-x_{1,k})^2+(s_{M,2}-x_{2,k})^2} \\ V_k\\ \phi_k
\end{array}
\right] \,,
\end{align*}
where we recall that $T$ is the sampling interval.

Consider the dynamical system of~\eqref{e:PCRLB_system}
and~\eqref{e:PCRLB_measurement}, and denote as $\P_k$ the covariance
matrix of any unbiased estimator of the system's state at time
$k$. Then, $\P_k$ is lower bounded by the inverse of the Fisher
information matrix $\J_k$, that is $ \J_k^{-1} \preceq \P_k$. Such a
matrix can be recursively computed as~\cite{Tichavsky1998}
\begin{align} \label{e:PCRLB}
\mathbf{J}_{k+1}= & \mathbf{D}_k^{22}-\mathbf{D}_k^{21} \left(
\mathbf{J}_k + \mathbf{D}_k^{11}\right)^{-1} \mathbf{D}_k^{12}
\end{align}
with
\begin{align}
\mathbf{D}_k^{11} = & -\E\left\{ \nabla_{\x_k} \left[ \nabla_{\x_k}
  \log p\left( \x_{k+1} | \x_k \right) \right] ^T
\right\} \,,  \label{e:D11}\\ 
\mathbf{D}_k^{21} = &
\left(\mathbf{D}_k^{12}\right)^T = -\E\left\{ \nabla_{\x_k} \left[
  \nabla_{\x_{k+1}} \log p\left( \x_{k+1} | \x_k \right) \right] ^T
\right\} \,, \label{e:D12}\\ 
\mathbf{D}_k^{22} = & -\E\left\{
\nabla_{\x_{k+1}} \left[ \nabla_{\x_{k+1}} \log p\left( \x_{k+1} |
  \x_k \right) \right] ^T \right\} \nonumber \\
&  - \E\left\{\nabla_{\x_{k+1}} \left[
  \nabla_{\x_{k+1}} \log p\left( \y_{k+1} | \x_k \right) \right] ^T
\right\} \label{e:D22} \,,
\end{align}
where the initial condition is given by the information matrix $\J_0$,
which is computed based on the initial prior density
$p\left(\x_0\right)$.

The system model in~\eqref{e:PCRLB_system}
and~\eqref{e:PCRLB_measurement} has additive Gaussian noise.  Hence,
the terms in~\eqref{e:D11},~\eqref{e:D12} and~\eqref{e:D22} may be
expressed as (see~\cite[eq.~(34) -- (36)]{Tichavsky1998})
\begin{align}
\mathbf{D}_k^{11} & = \E \left\{ \left[ \nabla_{\p_k} \f (\p_k)
  \right] \mathbf{Q}^{-1} \left[ \nabla_{\p_k} \f (\p_k) \right]^T
\right\} \,, \label{e:PCRLB_AWGN1}\\ \mathbf{D}_k^{12} & = - \E
\left\{ \left[ \nabla_{\p_k} \f (\p_k) \right] \right\} \Q^{-1}
\,, \label{e:PCRLB_AWGN2} \\ \mathbf{D}_k^{22} & = \Q^{-1} + \E
\left\{ \left[ \nabla_{\p_k} \h (\p_k) \right] \mathbf{R_k}^{-1}
\left[ \nabla_{\p_k} \h (\p_k) \right]^T
\right\} \,. \label{e:PCRLB_AWGN3}
\end{align}
In general, there is no guarantee that the expectations in \eqref{e:PCRLB_AWGN1}-\eqref{e:PCRLB_AWGN3} exist. In  other words, the CRLB may not exist. Therefore, in the following we establish the existence of the CRLB by deriving closed form expressions for the first two
terms, namely $\mathbf{D}_k^{11}$ and $\mathbf{D}_k^{12}$, while we
derive bounds for the third term $\mathbf{D}_k^{22}$.
We start by proving the following useful Lemma:
\begin{lemma} \label{lemma:exp_trig}
Consider the dynamical system in \eqref{e:PCRLB_system} --
\eqref{e:PCRLB_measurement}, denote the speed and the orientation at
time $k=0$ as $V_0$ and $\phi_0$, respectively, and define the
following quantities: $ c_0 \triangleq \cos{\phi_0} ,\quad s_0
\triangleq \sin{\phi_0} ,\quad c_0' \triangleq \cos{2\phi_0} ,\quad
s_0' \triangleq \sin{2\phi_0} ,\quad \varepsilon_k \triangleq e^{-
  \frac{\left( k-1 \right) \sigma_4^2}{2}}$. Then
\begin{align*}
    \E \left\{ V_k \right\} & = V_0 \,, \qquad \E \left\{ V^2_k \right\} =
    \left( k-1 \right) \sigma_3^2 + V_0^2\\ \E \left\{ \cos{\phi_k}
    \right\} & = c_0 \varepsilon_k\,, \qquad  \E \left\{ \sin{\phi_k} \right\}
     = s_0 \varepsilon_k \\ \E \left\{ \sin{\phi_k} \cos{\phi_k}
    \right\} & = \frac{1}{2} s_0' \varepsilon_k^2 \,, \qquad \E \left\{
    \cos^2{\phi_k} \right\}  = \frac{1}{2} + \frac{1}{2} c_0'
    \varepsilon_k^4 
\end{align*}
\end{lemma}
\begin{IEEEproof}
See Appendix \ref{appendix_lemma_exp_trig}.
\end{IEEEproof}

Then, we note that the gradient of $ \f $ is easily evaluated as
\begin{align}
\nabla_{\p_k} \f (\p_k) = \left[
    \begin{array}[c]{cccc}
    1 & 0 & 0 & 0 \\ 0 & 1 & 0 & 0 \\ T \cos{\phi_k} & T \sin{\phi_k}
    & 1 & 0 \\ - T V_k\sin{\phi_k} & T V_k\cos{\phi_k} & 0 & 1
    \end{array} 
\right] \,. \label{e:gradient_f}
\end{align}
By substituting Eq.~\eqref{e:gradient_f} in Eq.\eqref{e:PCRLB_AWGN1}
and using Lemma~\ref{lemma:exp_trig} we obtain Eq.~\eqref{Dk11} (see top of next page). 
\begin{figure*}
\begingroup
\begin{align}
\mathbf{D}_k^{11} & = \E \left\{ \left[
    \begin{array}[c]{cccc}
    \sigma_1^{-2} & 0 & \frac{T }{\sigma_1^{2}}\cos{\phi_k} &
    -\frac{T}{\sigma_1^{2}} V_k \sin{\phi_k} \\ 0 & \sigma_2^{-2} &
    \frac{T }{\sigma_2^{2}}\sin{\phi_k} & \frac{T}{\sigma_2^{2}} V_k
    \cos{\phi_k} \\ \frac{T}{ \sigma_1^{2}} \cos{\phi_k} &
    \frac{T}{\sigma_2^{2}} \sin{\phi_k} & \frac{T^2}{\sigma_1^{2}}
    \cos^2{\phi_k} + \frac{T^2}{\sigma_2^{2}} \sin^2{\phi_k} +
    \sigma^{-2}_3 & T^2 \left( \frac{\sigma_1^{2} -
      \sigma_2^{2}}{\sigma_1^{2} \sigma_2^{2}} \right) V_k
    \cos{\phi_k} \sin{\phi_k}\\ - \frac{T}{\sigma_1^{2}} V_k
    \sin{\phi_k} & \frac{T}{\sigma_2^{2}} V_k \cos{\phi_k} & T^2
    \left( \frac{\sigma_1^{2} - \sigma_2^{2}}{\sigma_1^{2}
      \sigma_2^{2}} \right) V_k \cos{\phi_k} \sin{\phi_k} &
    \frac{T^2}{\sigma_1^{2}} V_k^2 \sin^2{\phi_k} +
    \frac{T^2}{\sigma_2^{2}} V_k^2 \cos^2{\phi_k} + \sigma_4^{-2}
    \end{array}
\right] \right\} \nonumber \\ & = \left[
    \begin{array}[c]{cccc}
    \sigma_1^{-2} & 0 & \frac{T c_0}{\sigma_1^{2}} \varepsilon_k & -
    \frac{T V_0 s_0}{\sigma_1^{2}} \varepsilon_k \\ 0 & \sigma_2^{-2}
    & \frac{T s_0}{\sigma_2^{2}} \varepsilon_k & \frac{T V_0
      c_0}{\sigma_2^{2}} \varepsilon_k \\ \frac{T c_0}{\sigma_1^{2}}
    \varepsilon_k & \frac{T s_0}{\sigma_2^{2}} \varepsilon_k & T^2
    \left( \frac{\sigma_1^{2} + \sigma_2^{2}}{2\sigma_1^{2}
      \sigma_2^{2}} + \frac{\sigma_2^{2} - \sigma_1^{2}}{2\sigma_1^{2}
      \sigma_2^{2}} c_0' \varepsilon_k^4 \right) + \sigma_3^{-2} & T^2
    \left( \frac{\sigma_1^{2} - \sigma_2^{2}}{\sigma_1^{2}
      \sigma_2^{2}} \right) \frac{V_0 s_0'}{2}
    \varepsilon_k^4\\ -\frac{T V_0 s_0}{\sigma_1^{2}} \varepsilon_k &
    \frac{T V_0 c_0}{\sigma_2^{2}} \varepsilon_k & T^2 \left(
    \frac{\sigma_1^{2} - \sigma_2^{2}}{\sigma_1^{2} \sigma_2^{2}}
    \right) \frac{V_0 s_0'}{2} \varepsilon_k^4 & T^2 \left( k -1
    \right) \sigma_3^{2} \left(\frac{\sigma_1^{2} +
      \sigma_2^{2}}{2\sigma_1^{2} \sigma_2^{2}} + \frac{\sigma_1^{2} -
      \sigma_2^{2}}{2\sigma_1^{2} \sigma_2^{2}} c_0' \varepsilon_k^4
    \right) + \sigma_4^{-2}
    \end{array}
\right] \,. \label{Dk11}
\end{align}
\endgroup 
\end{figure*}
Note that, when $k$ tends to infinity, this matrix tends to
a diagonal matrix, because $\varepsilon_k$ tends to zero, but the
right-lower element tends to infinity. This implies that, as $k$
grows, the most relevant contribution to the Fisher information matrix
in Eq.~\eqref{e:PCRLB} is given by the $\mathbf{D}_k^{22}$
term. Therefore, in the limit case, only the measurements influence
the PCRLB, not the model.

\begin{figure*}[th]
\begin{align}
\mathbf{\Pi} & = \E \left\{ \left[
\begin{array}[c]{cc}
\sigma_{r_1}^{-2}d^2_{1,1} + \ldots + \sigma_{r_M}^{-2}d^2_{1,M} &
\sigma_{r_1}^{-2}d_{1,1}d_{2,1} + \ldots +
\sigma_{r_M}^{-2}d_{1,M}d_{2,M} \\ \sigma_{r_1}^{-2}d_{1,1}d_{2,1} +
\ldots + \sigma_{r_M}^{-2}d_{1,M}d_{2,M} & \sigma_{r_1}^{-2}d^2_{2,1}
+ \ldots + \sigma_{r_M}^{-2}d^2_{2,M}
\end{array}
\right] \right\} \,. \label{PI}
\end{align}
\end{figure*}

Similarly, by substituting Eq.~\eqref{e:gradient_f} in
Eq.\eqref{e:PCRLB_AWGN2}, we obtain:
\begin{align*}
\mathbf{D}_k^{12} & = - \left[
    \begin{array}[c]{cccc}
    \sigma_1^{-2} & 0 & 0 & 0 \\ 0 & \sigma_2^{-2} & 0 & 0 \\ T
    \sigma_1^{-2} c_0 \varepsilon_k & T \sigma_2^{-2} s_0
    \varepsilon_k & \sigma_3^{-2} & 0 \\ - T \sigma_1^{-2} V_0 s_0
    \varepsilon_k & T \sigma_2^{-2} V_0 c_0 \varepsilon_k & 0 &
    \sigma_4^{-2}
    \end{array}
\right] \,.
\end{align*}

We now turn our attention to the $\mathbf{D}_k^{22}$ term in
\eqref{e:PCRLB}. For the sake of simple notation, we drop the $k$
subscript in the remainder of this section and we define the following
quantity:
\begin{align*}
d_{i,j} \triangleq \frac{x_i - s_{j,i}}{\sqrt{\left( x_1 - s_{j,1}
    \right)^2 + \left( x_2 - s_{j,2} \right)^2}} \,.
\end{align*}
After some algebra, the gradient of $\h$ is
\begin{align}
\nabla_{\p} \h (\p) & \triangleq
\left[
\begin{array}[c]{ccccc}
d_{1,1} & \ldots & d_{1,M} & 0 & 0 \\ d_{2,1} & \ldots & d_{2,M} & 0 &
0 \\ 0 & \ldots & 0 & 1 & 0 \\ 0 & \ldots & 0 & 0 & 1
\end{array}
\right] \,. \label{e:gradient_h}
\end{align}
Let us introduce the following definition for the matrix
$\mathbf{\Pi}$ in Eq.~\eqref{PI} (see top of next page).
By substituting Eq.~\eqref{e:gradient_h} in Eq.~\eqref{e:PCRLB_AWGN3},
and using the above definitions, we obtain
\begin{align} \label{e:D22_blocks}
\D^{22} & = \Q^{-1} + \left[
\begin{array}[c]{cc}
\mathbf{\Pi} & \mathbf{0} \\ \mathbf{0} & \mathbf{R}_2^{-1}
\end{array}
\right] \,,
\end{align}
where $\mathbf{0}$ is the $2 \times 2$ zero matrix and $\mathbf{R}_2$
is defined as the $2 \times 2$ right-lower block of the measurement
noise covariance matrix, that is $\mathbf{R}_2 \triangleq
\textrm{diag} \left( \sigma_V^2 \,, \, \sigma_{\phi}^2 \right)$.

In Eq.~\eqref{e:D22_blocks}, there are two main issues: 
first, one has to show that the expectations in the submatrix 
$\mathbf{\Pi}$ exist, and then one has to compute them. 
These expectations  are taken over non-linear functions of the state random
variables, which makes it hard to find a closed-form
expression of the PCRLB. This is a well known issue, as pointed out
in~\cite{Brehard06}, where after arguing the difficulty of deriving a
PCRLB in cartesian coordinates, as we do, the PCRLB has been derived
by using logarithmic polar coordinates. However, we note that the
model adopted in \cite{Brehard06} does not not include speed and
orientation measurements by on-board sensors, and that the derivation
of the PCRLB in a coordinate system does not give insight on which is
the best estimator on another coordinate system \cite{Boyd2004, Ristic_book}, which greatly limits the model of \cite{Brehard06} for our case. 
In the cartesian coordinate system one could resolve the issue by a Monte Carlo approach to calculate numerically the PCRLB. 
However, analytical expressions of the PCRLB are of fundamental interest for many localization applications, such as when trying to
understand where to place the slaves in order to maximize the information
content given by the Fisher information
matrix \cite{Bishop+10,Shames+11}, or when planning the path of a
robot to minimize the uncertainty in its location. 
This further motivates the derivation of the analytical results that we present below. 

We must first prove that the expectations in the submatrix 
$\mathbf{\Pi}$ exist. In the following, we establish the existence of such expectations by deriving them in closed-form, and we derive closed-form expressions of upper and lower bounds for the entries of $\mathbf{\Pi}$ for which the exact closed form expression is too complex for practical numerical evaluations.  Then, we calculate them numerically in Section~\ref{simulations}.
To compute the expectations on the diagonal elements of $\mathbf{\Pi}$, we
establish the following result.

\begin{proposition} \label{prop:closed_form_diag}
Let $q$ and $z$ be independent Gaussian random variables with average
and standard deviation $\mu_q$, $\sigma_q$ and $\mu_z$ and $\sigma_z$,
respectively, with $\sigma_z=\sigma_q$ \footnote{Recall that these are the variances of the process noise on the coordinates, see Eq.\eqref{e:PCRLB_system}.}. 
Then, 
\begin{align}
\E \left\{ \frac{\displaystyle q^2}{\displaystyle q^2+z^2}\right\} 
= \frac{1}{1+\Upsilon} 
\left[ 1 + \sum_{k=1}^\infty \left( -1 \right)^k \frac{\mu_{F,k}}
{\left( 1+\Upsilon \right)^k}\right] \,,
\end{align}
where 
$$
\Upsilon
\triangleq 
\frac{\displaystyle \mu_z^2 + \sigma_z^2}{\displaystyle \mu_q^2 + \sigma_q^2}
\left[ 1 + \sum_{k=1}^\infty \left( -1 \right)^k 
\frac{\sigma_q^{2k} \mu_k}{\left( \mu_q^2 + \sigma_q^2 \right)^k}
\right] \,,
$$
and $\mu_k$ denotes the $k$-th central moment of the noncentral chi-squared distribution of 1 degree of freedom, and $\mu_{F,k}$ denotes the $k$-th central moment of the doubly noncentral F-distribution.
\end{proposition}

\begin{IEEEproof}
See Appendix \ref{appendix_prop_closed_form_diag}.
\end{IEEEproof}

The previous proposition allows us to establish the existence of the expectations on the diagonal elements of $\mathbf{\Pi}$. A similar result can be derived for the off-diagonal elements. Unfortunately, these closed form expressions are of limited practical usage due to the computational complexity of the many summations and function evaluations therein involved. Accordingly, we derive the following upper and lower bounds, which are much more easy to use in practice:

\begin{proposition} \label{prop:diag}
Let $q$ and $z$ be independent Gaussian random variables of average
and standard deviation $\mu_q$, $\sigma_q$ and $\mu_z$ and $\sigma_z$,
respectively. Then
\begin{align*}
e^{\alpha - \ln{\left(\sigma_q^2 + \mu_q^2 + \sigma_z^2 +
    \sigma_z^2\right)}} \leq \E \left\{\frac{q^2}{q^2+z^2}\right\}
\leq 1 \,,
\end{align*}
where $ \alpha = - \gamma_e - \ln{\left(2\right)} -
2\ln{\left(\sigma_q\right)} - M\left( 0, \, 1 / 2, \,
-\mu_q^2 / 2\sigma_q^2 \right)\,, $ the symbol $\gamma_e$
denotes the Euler Gamma constant, and the symbol $M(a,b,z)$ denotes
Kummer's confluent hypergeometric function \cite{Abramowitz}.
\end{proposition}
\begin{IEEEproof}
See Appendix \ref{appendix_prop_diag}.
\end{IEEEproof}

%
%
To compute bounds on the off-diagonal elements of $\mathbf{\Pi}$, we
give the following proposition:
\begin{proposition} \label{prop:off_diag}
Let $q$ and $z$ be independent Gaussian random variables with average
and standard deviation $\mu_q$, $\sigma_q$, and $\mu_z$ and $\sigma_z$,
respectively. Let the expectation be computed in the Cauchy principal
value sense. Then
\begin{align*}
- \frac{1}{2} \leq \E\left\{ \frac{qz}{q^2+z^2}\right\} \leq
\frac{1}{2} \,,
\end{align*}
\end{proposition}
\begin{IEEEproof}
The bounds follow immediately from that $- 1/2 \leq
{qz}/(q^2+z^2) \leq 1/2, \, \forall q \in \R, \forall z
\in \R$.
\end{IEEEproof}

We are now in the position to compute the upper and lower bounds on
the Fisher information matrix, in the positive semidefinite sense, by
the following result.
\begin{proposition} \label{prop:ublbpd}
Let $\J^{(ub)}$ and $\J^{(lb)}$ be two matrices consisting of
element-wise upper and lower bounds, respectively, of the Fisher
information matrix $\J$ in Eq.~\eqref{e:PCRLB} as obtained by
Propositions~\ref{prop:diag}~and~\ref{prop:off_diag} for the elements
of $\mathbf{\Pi}$ of Eq.~\eqref{e:D22_blocks}. Let
\begin{align}
\J^{(ub, G)}_{ii} \triangleq & \left\{
\begin{array}{l}
    \J^{(ub)}_{ii}\,, \qquad \textrm{if} \quad \J^{(ub)}_{ii} > \min \left(
    u_i^{(row)} \,, \, u_i^{(col)}\right)\\ \min \left( u_i^{(row)}
    \,, \, u_i^{(col)}\right) + \epsilon \,, \\ \quad \textrm{if} \quad
    \J^{(ub)}_{ii} \leq \min \left( u_i^{(row)} \,, u_i^{(col)}\right)  \label{e:gershgorin_first}
\end{array}
\right. \\ 
\J^{(ub, G)}_{ij} \triangleq &
\J^{(ub)}_{ij} \,, \quad \forall \: i,j = 1 \ldots n \,, \,
\textrm{with} \: i \neq j \,,\\ \J^{(lb, G)}_{ii} \triangleq &
\J^{(lb)}_{ii} \,, \quad \forall \: i = 1 \ldots n \,, \\ \J^{(lb,
  G)}_{ij} \triangleq & \left\{
\begin{array}{l}
    \J^{(ub)}_{ij} \quad  \textrm{if} \quad \J^{(lb)}_{ii} > \min \left(
    l_i^{(row)} \,, \, l_i^{(col)}\right)\\ 
    \left( \min \left(\J^{(lb)}_{ii} / l_i^{(row)} \,, \, \J^{(lb)}_{ii} /
    l_i^{(col)}\right) - \epsilon \right) \J^{(lb)}_{ij}  \\ 
    \quad \textrm{if} \quad \J^{(lb)}_{ii} \leq \min \left( l_i^{(row)} \,, \,
    l_i^{(col)}\right)
\end{array}
\right. \nonumber \\ 
& \quad \forall \: i,j = 1 \ldots n \,, \,
\textrm{with} \: i \neq j \,, \label{e:gershgorin_last}
\end{align}
where $u_i^{(row)} = \displaystyle\sum\limits_{j \neq i}\left|
\J^{(ub)}_{ij} \right|$ , $u_i^{(col)} = \displaystyle\sum\limits_{j
  \neq i}\left| \J^{(ub)}_{ji} \right|$ , $l_i^{(row)} =
\displaystyle\sum\limits_{j \neq i}\left| \J^{(lb)}_{ij} \right|$ ,
$l_i^{(col)} = \displaystyle\sum\limits_{j \neq i}\left|
\J^{(lb)}_{ji} \right|$ and $\epsilon$ is an arbitrary constant with
$\epsilon > 0$. Then
\begin{align}
\mathbf{0} \preceq \J^{(lb, G)} \preceq \J \preceq \J^{(ub, G)} \,.
\end{align}
\end{proposition}
\begin{IEEEproof}
See Appendix \ref{appendix_prop_ublbpd}.
\end{IEEEproof}

This concludes the steps to establish the existence and upper and lower bounds of the Fisher information matrix. In the following section, we use these results to compare the performance of our proposed localization method to the CRLB.

\section{Simulation Results}
\label{simulations}

In this section we present simulation results to validate both the
PCRLB derivations and the estimator we presented in the previous
sections. In particular, we compare our new estimator based on
Pareto optimization that we have developed in Section~\ref{derivation}
to some solutions from the literature and to the PCRLB we
derived in Section~\ref{section:CRLB}.

\subsection{Performance of the proposed sensor fusion method}

\begin{figure}[th]
\centering 
\includegraphics[width=0.99\columnwidth, trim=25 10 30 20]{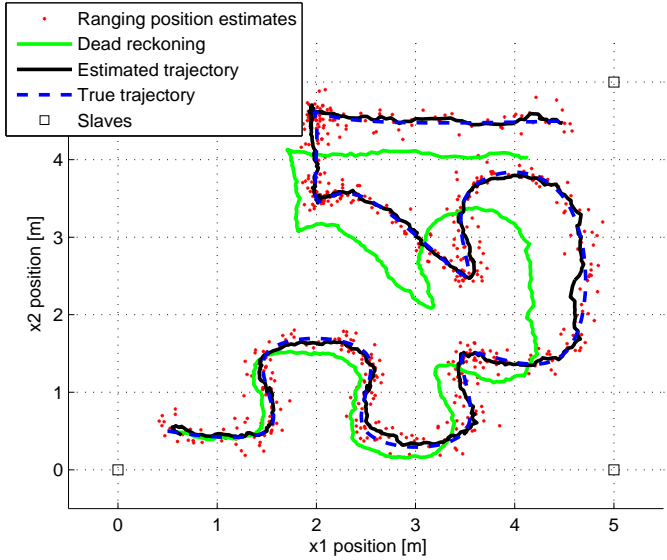}
\caption{PWL-acceleration trajectory case, scenario B. The proposed
  sensor fusion technique closely tracks the true trajectory with RMSE
  = 5.5 cm.} \label{F:ParetoWLS_traj_spline}
\end{figure}

In this section we present extensive simulation results of our new
estimator~\eqref{e:model_old} based on the
optimal fusion coefficient of Eq.~\eqref{e:problem2_solution} and
Pareto weighting factor of Eq.~\eqref{e:Pareto_rho_selection}.  The
values of the parameters used in the numerical simulations are the
following:
\begin{itemize}
\item Speed measurement noise standard deviation $\sigma_v = 0.05\:
  $m/s, to model the worst-case performance of an odometry sensor,
  \cite{Mourikis2006};
\item Orientation measurement noise standard
  deviation $\sigma_{\phi} = \pi/8\:$~rad, to model the worst-case
  performance of a magnetometer subject to disturbances due to the
  environment, see, e.g., \cite{Georgiev2004},
  \cite{Abbott1999};
\item Ranging noise model parameters (in
  Eq.~\eqref{e:range_variance_model}): $\sigma_0=0.25$,
  $\kappa_{\sigma}=0.25$;
\item Sampling time interval $T = 0.1\: $s.
\end{itemize}
The following two scenarios have been considered:
\begin{enumerate}[(A)]
\item Linear trajectory with constant speed of 0.1 m/s. This is a
  deterministic trajectory, that is, no process noise is added. As an example, this scenario emulates the real-world situation in which the mobile node is on a vehicle moving along a straight line without performing any maneuver.
\item Piece-wise-linear (PWL)-acceleration trajectory, which is more
  complex and realistic than the linear one, with a maximum
  acceleration of 0.5 m/s$^2$, see Fig.~\ref{F:ParetoWLS_traj_spline}. This scenario emulates a mobile node performing maneuvers with an acceleration range consistent with the indoor environment, e.g. a walking person carrying the mobile node.
\end{enumerate}

The absolute value of $\beta^*$, provided by the algorithm, has been
clipped to 0.99 to ensure that the average of the estimation error is
non-expansive, as discussed in Section~\ref{derivation}.

Note that, in \ref{Proposition_5}, in the expression for $\gamma$ the true speed $V_k$ and
orientation $\phi_k$ terms are present. Since these are not available,
we use these approximations:
\begin{align}
V_{\rm approx} \left( k \right) \simeq & \frac{1}{T} \left( \left(
\hat{x}_{1,k \mid k} - \hat{x}_{1,k-1 \mid k-1} \right)^2 \right. \nonumber \\
& \left. + \left(\hat{x}_{2,k \mid k} - \hat{x}_{2,k-1 \mid k-1} \right)^2
\right)^{\frac{1}{2}} \,, \label{e:approx_v}\\ 
\phi_{\rm approx} \left( k \right) \simeq
& \arctan \left( \frac{ \hat{x}_{2,k \mid k} - \hat{x}_{2,k-1 \mid
    k-1} } { \hat{x}_{1,k \mid k} - \hat{x}_{1,k-1 \mid k-1} } \right)
\,.\label{e:approx_phi}
\end{align}
Also, note that the second-order moment of the ranging error, given by
\eqref{e:somr}, depends on the true ranges $r_i$.  Therefore, for the
calculation of the term
$\E\left\{\left(w_{x_{1,k+1}}^{(r)}\right)^2\right\}$ in $b_k$ and
$a_k$, we use the approximations for the ranges
\begin{align}
r_{i,k+1}^{(\rm
  approx)}=\sqrt{\left(s_{i,1}-\hat{x}_k\right)^2+\left(s_{i,2}-\hat{y}_k\right)^2}
\,, \quad i = 1, \ldots , M \,. \label{e:approx_r}
\end{align}

\begin{figure}[th]
\centering 
\includegraphics[width=0.99\linewidth, trim=20 10 30 20
  ,clip=]{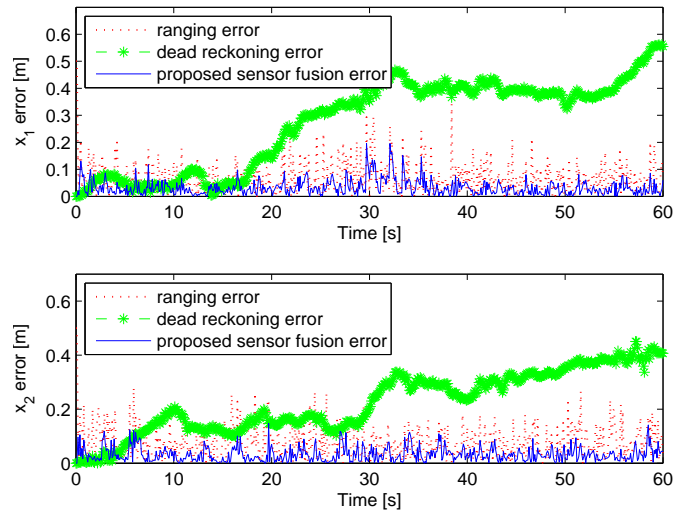}
\caption{Absolute errors in scenario A.  The dead-reckoning estimates,
  asterisks, have a high bias, while the WLS ranging estimates, dashed
  line, present a higher variance, but a reduced bias. The proposed
  sensor fusion technique is able to reach a good tradeoff between
  estimator bias and variance.} \label{F:Pareto_error_line}
\end{figure}

\begin{figure}[th]
\centering 
\includegraphics[width=0.99\linewidth]{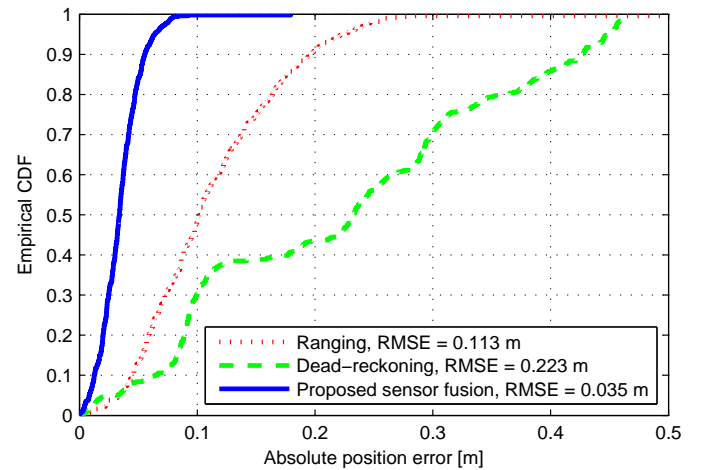}
\caption{Empirical CDF of the absolute positioning error in 2D for
  scenario A. Using the proposed method (blue line), about 95\% of the
  position estimates have an error smaller than 7 cm, and the
  performance is improved with respect to both the ranging and the
  dead-reckoning stand-alone systems.} \label{F:ParetoWLS_cdf_line}
\end{figure}
\begin{figure}[th]
\centering \small
\begin{tabular} {c c}
\multicolumn{2}{c}{\textsc{Table I}}
\\ \multicolumn{2}{c}{\textsc{Average Execution Time Per Iteration}
  [s].} \\ \multicolumn{2}{c}{\footnotesize Matlab$^\circledR$
  implementation on an Intel$^\circledR$ Core2 Duo,}
\\ \multicolumn{2}{c}{\footnotesize 2.40 GHz PC with 3 GB RAM.}
\\ \hline \emph{EKF} & $1.3 \times 10^{-4}$ \\ \hline \emph{MSE}
Eq.~\eqref{e:special_case_sol} & $1.7 \times 10^{-4}$ \\ \hline
\emph{LC-KF} & $1.9 \times 10^{-4}$ \\ \hline \emph{UKF} & $4.2 \times
10^{-4}$ \\ \hline Proposed sensor fusion, Eq.~\eqref{e:problem2_solution}
& $4.5 \times 10^{-4}$ \\ \hline \hline
\end{tabular}
\end{figure}

The absolute error and empirical Cumulative Distribution Function
(CDF) obtained by our new method for scenario A are shown in
Fig.~\ref{F:Pareto_error_line}~and~\ref{F:ParetoWLS_cdf_line}.  In
this scenario, a root-mean-squared error (RMSE) of 4.0 cm in the
position estimate over the entire trajectory has been obtained by our
method.
Furthermore, in Fig.~\ref{F:ParetoWLS_traj_spline} the more realistic
PWL-acceleration case is shown, where an RMSE of 5.5 cm is obtained.

Numerical simulation results show that the approximations \eqref{e:approx_v}-\eqref{e:approx_r} give good
performance and that the proposed method yields a
good trade-off between variance and bias of the estimated position. 
Therefore, the proposed estimator overcomes the limitation
of dead reckoning (that is the slowly accumulating bias) while also
reducing the relatively higher variance of the ranging estimator. The
result is an overall smooth and accurate estimate of the
trajectory. In the following, we compare the proposed estimator to
other solutions that are commonly adopted in similar localization
problems.

\subsection{Performance Comparison}

Here we compare the sensor fusion based on Pareto optimization proposed in this paper
to several other methods in the two mobility scenarios considered in
the previous section. To provide an extensive comparison, the
simulations are performed for different values of the sampling period
$T$ and of the speed $V$ for scenario A, whereas
for scenario B the simulations are
performed for different values of $T$ and of the maximum
acceleration. Moreover, for each configuration, the simulations are
performed for 10 realizations of the random noises and the resulting
RMSE values are averaged.

The sensor fusion methods available from the literature that we have
considered are the following:
\begin{itemize}
\item Extended Kalman Filter (EKF)~\cite{Kailath}, where the measured
  speed and orientation data are employed as inputs, and the ranges as
  measurements. No motion model is assumed.
 \item Unscented Kalman Filter (UKF)~\cite{Julier2004}, with the same
   state-space model as the EKF above.
\item Loosely Coupled Kalman Filter (LC-KF), where range measurements
  are preprocessed with the WLS algorithm to obtain preliminary
  position measurements, thus using a linear measurement equation. The
  usual linearization of the state function is performed for
  calculating the input noise matrix \cite{Kailath}.
\end{itemize}
Notice that the methods based on the Kalman filter theory have been
implemented without assumptions about the motion model, to provide a
fair comparison with the developed method of this paper.  If prior
information about motion models is included in the state space
formulation, the Kalman approaches might be able to provide better
performance, as we will show in the example in the next
section. Notice also that we provide results of the MSE special case
when $\rho_{1,k}=0.5$ $\forall k$ in
Proposition~\ref{Proposition_5}. This method does not provide good
performance compared to the more general case of choosing the best
$\rho_{1,k}$ by problem~\eqref{e:Pareto_rho_selection}. However, it
features a much smaller computational complexity.

\begin{figure}
\centering 
\includegraphics[width=0.99\linewidth, trim=20 0 35 10
  ,clip=]{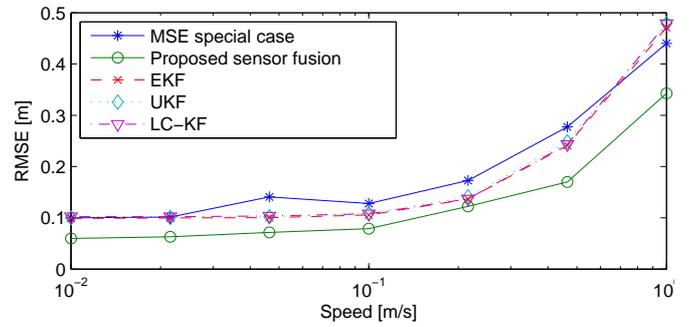}
\caption{RMSE vs speed of the different methods for Scenario A with
  T~=~0.5~s. }
\label{F:comparison_linear_0p5s}
\end{figure}

\begin{figure}
\centering \includegraphics[width=0.99\linewidth, trim=18 0 35 10
  ,clip=]{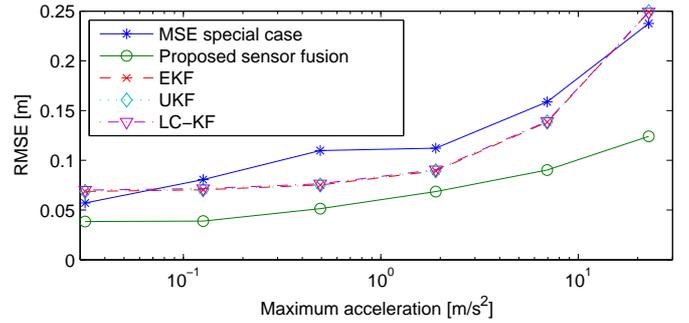}
\caption{RMSE vs maximum acceleration of several methods for Scenario
  B with T = 0.1 s. } \label{F:comparison_spline_0p1s}
\end{figure}

Comparisons for two selected simulation configurations are provided in
Fig.~\ref{F:comparison_linear_0p5s} and
\ref{F:comparison_spline_0p1s}, as obtained in the two scenarios with
different values of $T$. The proposed sensor fusion based on Pareto optimization outperforms the
other methods, while the KF-based methods have similar performance
among them. However, the proposed sensor fusion based on Pareto optimization
requires the highest computational load among the implemented
techniques, as shown in the average simulation execution times listed
in Table~I, although it is close to the execution time of UKF. 

\subsection{CRLB evaluation}

In this section, we provide a comparison of the RMSE performance of
the proposed sensor fusion method based on Pareto optimization and the Kalman filter based methods we used in the previous section to the fundamental
limits given by the Cram\'er Rao lower bound. In particular, we first
provide comparisons with the \emph{Par}CRLB, and then with the PCRLB
as derived in Section~\ref{section:CRLB}.

\begin{figure*}
\centering 
\subfigure{ \includegraphics[width=0.80\textwidth]
  {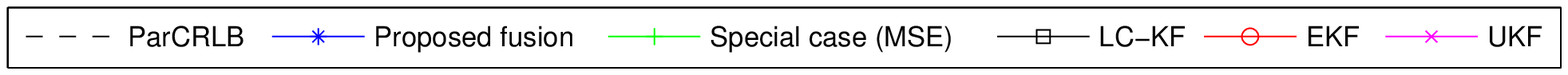}
}
\\  
\vskip -0.2cm
\setcounter{subfigure}{0}
\subfigure{ \includegraphics[width=0.80\textwidth]
  {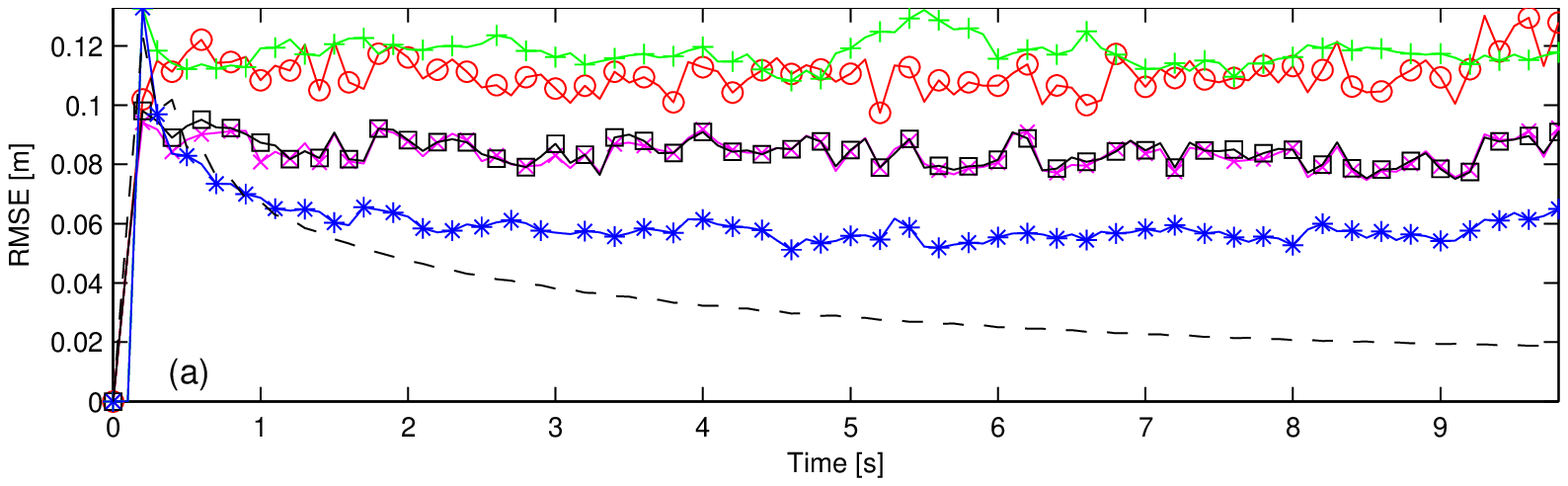}
   \label{F:CRLB_comparison_line}
}
\\
\subfigure{ \includegraphics[width=0.80\textwidth] {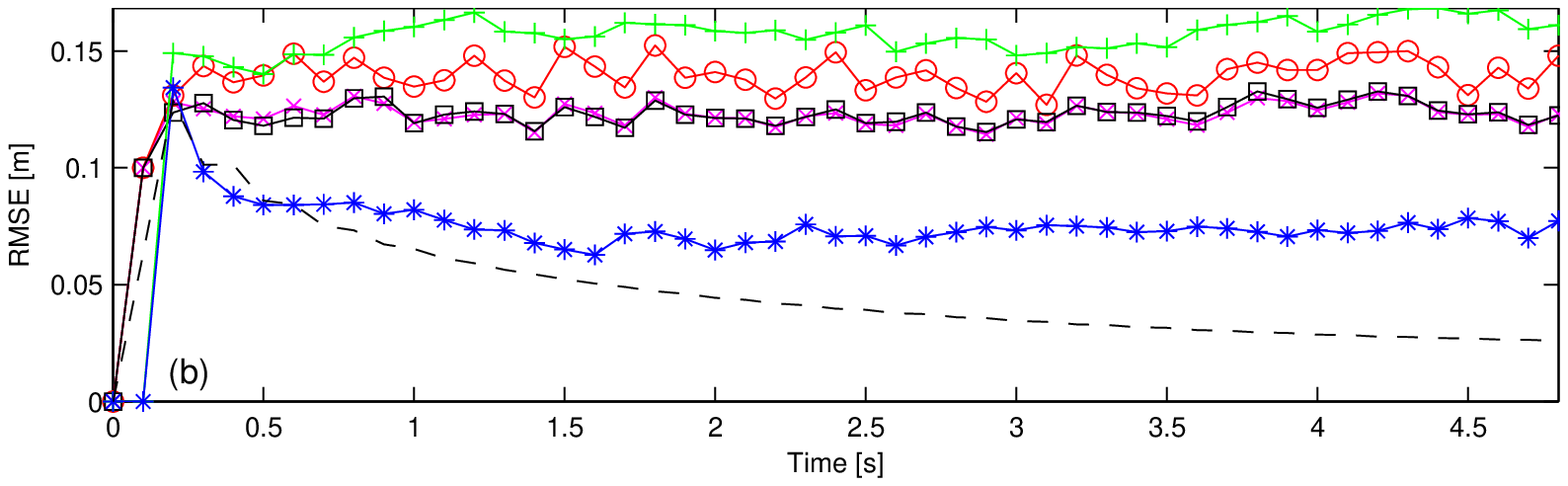}
   \label{F:CRLB_comparison_line_fast}
} 
\\ 
\subfigure{ \includegraphics[width=0.80\textwidth]
     {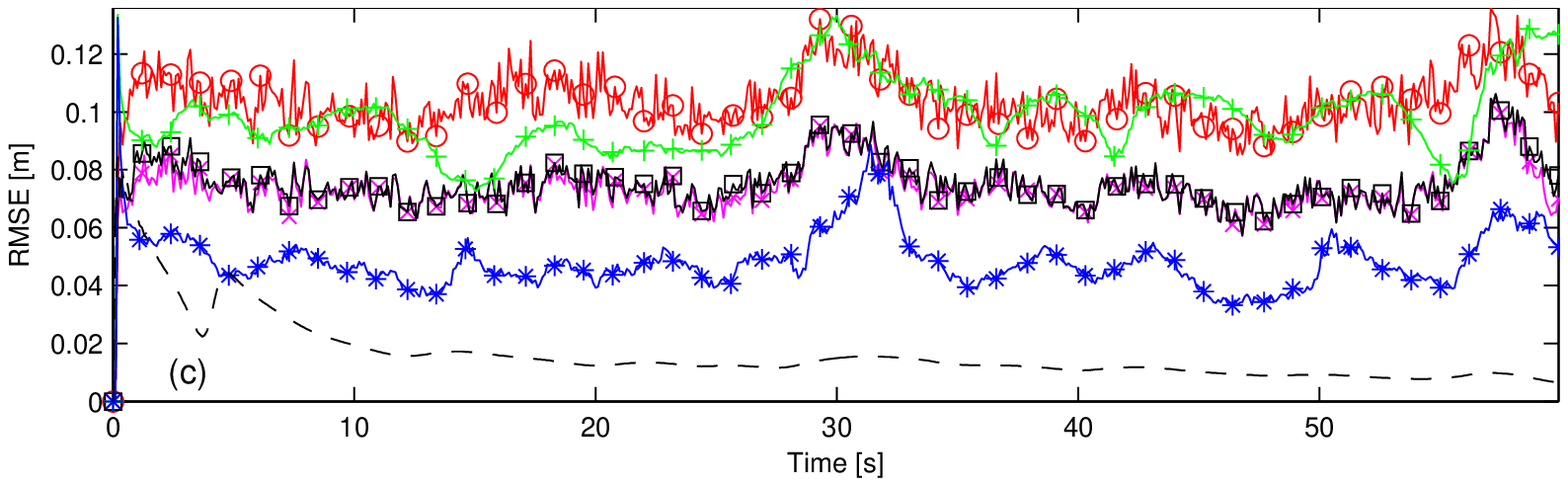}
   \label{F:CRLB_comparison_spline}
}
\\
\subfigure{ \includegraphics[width=0.80\textwidth] {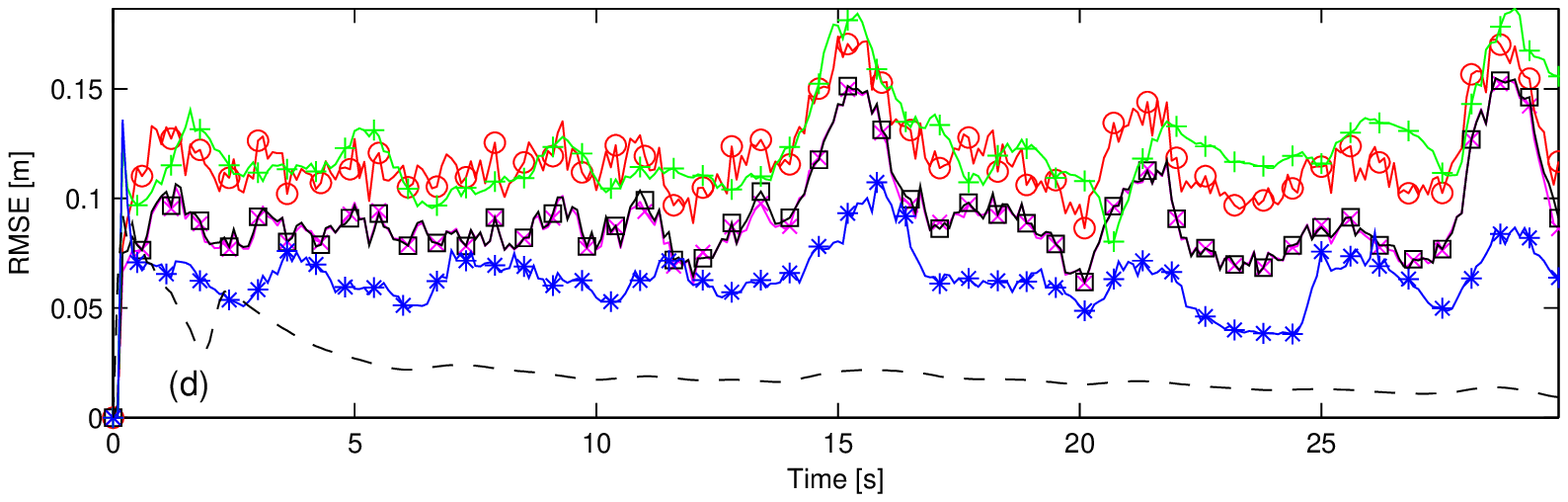}
   \label{F:CRLB_comparison_spline_fast}
}
\caption{Error performance of the proposed sensor fusion method based on Pareto optimization of Eq.~\eqref{e:problem2_solution} and other methods against the
  theoretical bound given by the square root of the trace of the
  \emph{Par}CRLB (dashed line): (a) linear trajectory, constant
  velocity of 0.5 [m/s]; (b) linear trajectory, constant velocity of 1
  [m/s]; (c) spline trajectory, maximum acceleration of
  0.3~$[\mbox{m/s}^2]$; (d) spline, maximum acceleration of
  1~$[\mbox{m/s}^2]$.} \label{F:CRLB_comparison}
\end{figure*}

\begin{figure}
\centering
\includegraphics[width = 0.90\linewidth]{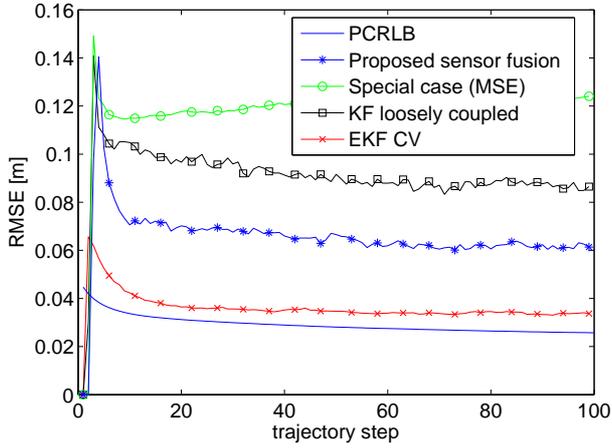}
\caption{
Numerical simulation results on the class of trajectories defined by the model in \eqref{e:PCRLB_system}-\eqref{e:PCRLB_measurement}. The behavior of the proposed sensor fusion method based on Pareto optimization is compared to other estimators from the literature and to the PCRLB derived in Section \ref{section:CRLB}. Error curves are obtained as the average of 1000 Monte Carlo iterations.
}
\label{f:PCRLB_MC}
\end{figure}

The performance of the filters was compared based on the RMSE of the
position estimate, where the error is defined as the Euclidean distance between the true position and the estimate. The results compared to the \emph{Par}CRLB for several specific trajectories are presented in
Fig.~\ref{F:CRLB_comparison}, where the error curves were obtained by
averaging over 100 Monte-Carlo runs. We notice that the proposed sensor fusion method based on Pareto optimization provides better performance than the other considered KF methods as it is the closest to the \emph{Par}CRLB. We stress that none of the considered methods makes assumptions about the motion model, therefore in this sense the \emph{Par}CRLB is an overly optimistic bound.
\\
\indent Furthermore, the numerical simulation results related to the model in \eqref{e:PCRLB_system}-\eqref{e:PCRLB_measurement} are shown in Fig. \ref{f:PCRLB_MC}, in comparison with the derived PCRLB in \eqref{e:PCRLB}. The method denoted as ``EKF CV'' is implemented using the motion model in \eqref{e:PCRLB_system}, and the same model is used to generate the trajectories. 
Specifically, a standard EKF is implemented \cite{Kailath}, by linearizing about the current estimate of the mean and covariance, where the Jacobian matrices are computed by numerical differentiation at each iteration.
In this case, we notice that EKF CV outperforms the proposed sensor fusion method based on Pareto optimization and approaches the PCRLB. This confirms that the knowledge of the specific underlying motion model is beneficial for estimation. However, the proposed sensor fusion method still provides good performance, improving over the LC-KF. Conversely to EKF CV, the proposed method does not assume any specific motion model, thus it can be applied to a wider class of trajectories, proving to be robust with respect to model mismatch. Nevertheless, the same Pareto optimization approach presented in this paper could be applied to derive an estimator which exploits the knowledge of the motion model, although such derivation is outside of the scope of this paper. 


\section{Conclusion} \label{conclusion}

In this paper, the fundamental limitations of position estimation for
a mobile node by the fusion of information from ranging, velocity and
angular measurements were investigated. Upper and lower bounds of the
posterior Cram\'er-Rao lower bound were derived. A sensor fusion
method based on Pareto optimization was proposed. The analysis was
validated by Monte Carlo simulations, which showed that the proposed
method provides better performance than Kalman filtering-based methods
when no knowledge of the dynamic model of the underlying system is
assumed or when the model is time varying.

Future work includes the extension of the proposed estimator to cases
when the mobility model is known.
The good performance achieved by the proposed sensor fusion method based on Pareto optimization in the case of no motion model assumption lets us think that the use of a motion model would substantially farther improve the performance of the proposed method.

\appendices
\section{Proof of Lemma \ref{Proposition_mean_ranging}}
\label{appendix_mean_ranging}
By taking the mean of Eq.~\eqref{e:positioning_estimate}, we obtain:
\begin{align*}
\E \left\{\hat{\x}^{(r)} \right\} = &
{\left(\A^T\W\A\right)}^{-1}\A^T\W  \notag \\
& \times \left(\1 \E \{ \tilde{r}_M^2\} - \E\left\{ \left[ \tilde{r}_1^2, \hdots , \tilde{r}_{M-1}^2 \right] ^T
\right\}+ \a \right) \notag \\
= & \x + \left(\A^T\W\A\right)^{-1}\A^T\W \notag \\
& \times \left( \1 \sigma^2_{w_{r_M}}
- \left[ \sigma^2_{w_{r_1}}, \hdots, \sigma^2_{w_{r_{M-1}}} \right] ^T
\right) \notag \\
= & \x + \E\left\{ \w_r \right\}\,,
\end{align*}
whereby the lemma follows.

\section{Proof of Lemma \ref{Proposition_correlation_ranging}}
\label{appendix_correlation_ranging}
By taking the square of Eq.~\eqref{e:range_expression} we get $
\tilde{r}^2_{i} = r^2_{i} + w^2_{r_i} + 2 r_{i} w_{r_i}\,. $ Then,
using this expression in \eqref{e:WLS_error_w_s} and substituting in
\eqref{e:positioning_estimate}, we have the following expression for
the ranging-based estimate:
\begin{align*}
\hat{\x}^{(r)} = & {\left(\A^T\W\A\right)}^{-1}\A^T \W \left\{ \left[
\begin{array}{c}
  r_M^2 - r_1^2, \ldots, r_M^2 - r_{M-1}^2
\end{array}
\right]^T \right. \\
& \left. + \B_k + \a \right\},
\end{align*}
where 
$$ \B_k = \left[
        \begin{array}{c}
            w_{r_M}^2 + 2 r_M w_{r_M} - w_{r_1}^2 - 2 r_1 w_{r_1}\\
            \vdots \\
            w_{r_M}^2 + 2 r_M w_{r_M} - w_{r_{M-1}}^2 - 2
            r_{M-1} w_{r_{M-1}}
        \end{array}
    \right]\,.  
$$

Therefore the error can be expressed as $ \w_r =
{\left(\A^T\W\A\right)}^{-1}\A^T\W \B_k\,, $ and its correlation is $
\E\left\{\w_r^2\right\} = {\left(\A^T\W\A\right)}^{-1}\A^T\W
\E\left\{\B_k \B^T_k\right\} \W^T\A{\left(\A^T\W\A\right)}^{-1}. $ We
let $\C \triangleq \E\left\{\B_k \B^T_k\right\}$, whereas the lemma
follows after some algebraic manipulation.

\section{Proof of Lemma \ref{Proposition_1}}
\label{appendix_Proposition_1}
Since $\tilde{V}_k$ and $\tilde{\phi} _k$ are independent,
\begin{align}
\E \left\{ \tilde{V}_k \cos\left(\tilde{\phi}_k\right) \right\} = 
& \E\left\{ \tilde{V}_k \right\} \E \left\{
\cos\left(\tilde{\phi}_k\right) \right\} \notag \\
= & V_k \E \left\{ \cos(\phi _k
+ w_{\phi_k}) \right\} \nonumber \\
= & V_k \left( \cos(\phi _k) \E \left\{ \cos( w_{\phi,k}) \right\}
\right.
\notag \\
& \left. - \sin(\phi _k) \E \left\{ \sin( w_{\phi,k}) \right\} \right) \,.
    \label{e:expectation_movement}
\end{align}
By using the cosine series,
\begin{align}
\MoveEqLeft \E \left\{ \cos( w_{\phi,k}) \right\} \hskip-0.5mm =  
\hskip-0.5mm \E \hskip-0.5mm \left\{ \hskip-0.5mm 1 - \frac{w^2_{\phi,k}}{2} + \frac{w^4_{\phi,k}}{4!}
+ \cdots + (-1)^n \frac{w^{2n}_{\phi,k}}{(2n)!} \hskip-0.5mm \right\}  \nonumber \\
& =  1 - \frac{\sigma_{\phi}^2 }{2} + \frac{\sigma_{\phi}^4 }{2^2 \cdot 2!}
+ \cdots + (-1)^n \frac{\sigma_{\phi}^{2n}}{2^n \cdot n! } =
e^{-\frac{\sigma_{\phi}^2}{2}} \,,
\label{e:cosine_expansion}
\end{align}
where we have used that $\tilde{\phi} _k$ is Gaussian. Similarly, by using the sine series, it results
\begin{align}
\E \left\{ \sin( w_{\phi,k}) \right\} = & \E \left\{ w_{\phi} -
\frac{w^3_{\phi,k}}{3!} + \frac{w^5_{\phi,k}}{5!} - \frac{w^7_{\phi,k}}{7!}
    \right. \notag \\
    & \left.
    + \cdots + (-1)^n \frac{w^{2n+1}_{\phi_k}}{(2n+1)!} \right\} = 0
    \,.
\label{e:sine_expansion}
\end{align}
Then Eq.~\eqref{e:expectation_inertial} follows by substituting
Eqs.~\eqref{e:cosine_expansion} and~\eqref{e:sine_expansion} in
Eq.~\eqref{e:expectation_movement}. Furthermore, since $\tilde{V}_k$ and $\tilde{\phi} _k$ are
independent, we have
\begin{align}
\E \left\{ \tilde{V}^2_k \cos^2(\tilde{\phi}_k) \right\} = & \E
\left\{ \tilde{V}^2_k \right\} \E \left\{ \cos^2(\tilde{\phi}_k)
\right\} \,.
\label{e:dead_reck_square}
\end{align}
For the second factor of the right-end side of this expression we have
\begin{align}
\E\left\{\cos^2(\tilde{\phi}_k)\right\} = & \E \left\{
  \frac{1}{2} + \frac{1}{2} \cos(2\tilde{\phi}_k) \right\}
\nonumber \\
= & 
\frac{1}{2} + \frac{1}{2} \E \left\{ \cos\left(2 \left( \phi_k +
w_{\phi,k} \right) \right) \right\} \nonumber \\ 
= & \frac{1}{2} + \frac{1}{2} \E \left\{ \cos\left(2\phi_k\right)
\cos\left(2w_{\phi,k}\right)
    \right. \nonumber \\
    & \left.
    - \sin\left(2\phi_k\right) \sin\left(2w_{\phi,k}\right) \right\}
\nonumber \\
= &
\frac{1}{2} + \frac{1}{2} \cos\left(2\phi_k\right) \E \left\{
\cos\left(2w_{\phi,k}\right) \right\} \,. \label{e:cos_squared}
\end{align}
Using the cosine series, we write
\begin{align}
\E \left\{ \cos\left(2w_{\phi,k}\right) \right\} = & \E \left\{
1-2^2\frac{{w^{2}_{\phi_k}}}{2!}  +2^4\frac{3{w^{4}_{\phi_k}}}{4!}
    \right. \nonumber \\
    & \left.
    -2^6\frac{5{w^{6}_{\phi_k}}}{6!}  + \cdots + (-1)^n
    \frac{(2w_\phi)^{2n}}{(2n)!}  \right\} \nonumber \\ = &
    1-2\frac{\sigma_\phi^{2}}{1!} + 2^2\frac{\sigma_\phi^{4}}{2!} -
    2^3\frac{\sigma_\phi^{6}}{3!}
    \nonumber \\
    &
    + \cdots + (-1)^n \frac{2^n \sigma^{2n}_\phi}{n!} =
    e^{-2\sigma^2_\phi} \,. \label{e:cos_double}
\end{align}
By substituting~\eqref{e:cos_double} in~\eqref{e:cos_squared}, we
obtain:
\begin{align}
\E\left\{\cos^2(\tilde{\phi}_k)\right\} = \frac{1}{2} + \frac{1}{2}
\cos\left(2\phi_k\right)e^{-2\sigma^2_\phi}\,.
\label{e:cos_squared_var}
\end{align}
Eq.~\eqref{e:2nd_order_inertial} follows by
substituting~\eqref{e:cos_squared_var} in~\eqref{e:dead_reck_square},
therefore concluding the proof.

\section{Proof of Lemma \ref{Proposition_3}}
\label{appendix_Proposition_3}
The expectation of~\eqref{e:model_old} is
\begin{align*}
\lefteqn{
\E \{ \hat{x}_{1,k+1 \mid k+1} \} = \left( 1
- \beta_{1,k} \right) x_{k + 1} + \left( 1 - \beta_{1,k} \right) \E
\left\{ w_{x_{1,k + 1}}^{(r)} \right\} 
} & 
\notag \\
 & \quad + \beta_{1,k} \left( x_k + \E\{w_{x_{1,k}}\} + T V_k \cos\left(\phi_k\right)
e^{-\frac{\sigma_{\phi}^2}{2}} \right) \notag \\
& = \left( 1 - \beta_{1,k} \right) x_{k + 1} + \beta_{1,k} x_{k +
      1} + \left( 1 - \beta_{1,k} \right) \E \left\{ w_{x_{1,k +
        1}}^{(r)} \right\} \notag \\
& \quad + \beta_{1,k} \E \{w_{x_{1,k}}\}
+ \beta_{1,k} T V_k \cos\left(\phi_k\right) \left(
    e^{-\frac{\sigma_{\phi}^2}{2}} - 1 \right)
\notag \\
& = x_{k + 1} + \E \{w_{x_{1,k+1}}\}.
\end{align*}
whereby the lemma follows.

\section{Proof of Proposition \ref{Proposition_5}}
\label{appendix_problem2_solution}
The cost function in \eqref{e:problem2_rho_simple} is always positive for any choice of $\beta_{1,k}
\in \R$ and of $\rho_{1,k}$, and it is a parabola in $\beta_{1,k}$,
which is due to that $a_k>0$. It follows that the cost function is
convex. Therefore, the optimal solution is given by computing the
derivative, and observing that the optimal solution must lie in the
interval $\mathscr{B}=[-1,1]$.

\section{Proof of Lemma \ref{lemma:exp_trig}}
\label{appendix_lemma_exp_trig}
We note that the speed at time $k$ can be written as $ V_k = V_0 +
v_{3,1} + \ldots + v_{3,k-1} \,.  $ Since the process noise in the
speed, $v_3$, is white, it follows that $V_k \sim \mathcal{N}\left(
V_0, \left( k-1 \right) \sigma_3^2 \right) $.  Similarly, for the
orientation $\phi_k$, we have that $\phi_k \sim \mathcal{N}\left(
\phi_0, \left( k-1 \right) \sigma_4^2 \right) $.  Then the Lemma
follows by using Lemma~\ref{Proposition_1}, after simple algebraic and
trigonometric calculations.

\section{Proof of Proposition \ref{prop:closed_form_diag}}
\label{appendix_prop_closed_form_diag}
First, we start by proving the following initial lemma:
\begin{lemma} \label{lemma:ratio}
Let $a$, $b$ be two independent random variables with non-zero average. Suppose that 
\begin{align} \label{conv_condition}
\lim_{k \rightarrow \infty }\frac{\E \{ \left(b-\E\left\{b\right\}\right)^k\}}{\left(\E\left\{b\right\}\right)^k} =0 \,.
\end{align}
Then,
\begin{align*}
\E\left\{ \frac{a}{b} \right\} & = 
\frac{\E \left\{a\right\}}{\E \left\{b\right\}} 
\left[ 1 + \sum_{k=1}^{\infty}\left( -1 \right)^k \frac{\E\left\{\left(b-\E\left\{b\right\}\right)^k \right\}}{\left(\E\left\{b\right\}\right)^k} \right]
\end{align*}
\end{lemma}
\begin{IEEEproof}
Since $a$ and $b$ are independent, we have that
\begin{align}\label{e:lemma_start}
\E\left\{ \frac{a}{b} \right\} = \E\left\{a\right\}\E\left\{\frac{1}{b}\right\} \,.
\end{align}
Furthermore, 
\begin{align}\label{e:lemma_interm_result1}
\E\left\{ \frac{1}{b} \right\} = \E\left\{b^{-1}\right\} = 
\E\left\{ \left( 1 + \frac{b-\E\left\{b\right\}}{\E\left\{b\right\}} \right)^{-1} \right\} \frac{1}{\E\left\{b\right\}} \,.
\end{align}
By using Taylor series expansion as in~\cite[Section 5-4]{Papoulis}, we can write
\begin{align}\label{e:lemma_interm_result2}
\E \hskip-0.5mm \left\{\hskip-0.5mm \left(\hskip-0.5mm  1 + \frac{b-\E\left\{b \right\}}{\E\left\{b\right\}} \hskip-0.5mm \right)^{-1}\hskip-0.5mm\right\} \hskip-0.5mm = \hskip-0.5mm 1 \hskip-0.5mm + \hskip-0.5mm \sum_{k=1}^{\infty}\left( -1 \right)^k \frac{\E \{ \left(b-\E\left\{b\right\}\right)^k\}}{\left(\E\left\{b\right\}\right)^k}
\end{align}
The sum in the previous equation converges as a consequence of assumption~\eqref{conv_condition}. Finally, by substituting \eqref{e:lemma_interm_result2} in \eqref{e:lemma_interm_result1} and \eqref{e:lemma_start}, the lemma follows.
\end{IEEEproof}

Now, using Lemma~\ref{lemma:ratio} above, we write the expectation as 
\begin{align} \label{e:closed_form_diag_init}
\lefteqn{\E \left\{ \frac{\displaystyle q^2}{\displaystyle q^2+z^2}\right\} 
= \E \left\{ \frac{1}{\displaystyle 1+\frac{z^2}{q^2}}\right\}
= \frac{1}{1+\E\left\{ \frac{\displaystyle z^2}{\displaystyle q^2} \right\}} 
} & \notag \\
& \quad \times
\left[ 1 + \sum_{k=1}^\infty \left( -1 \right)^k \frac{\displaystyle  \E \left\{ \left( \frac{\displaystyle z^2}{\displaystyle q^2} - \E\left\{ \frac{\displaystyle z^2}{\displaystyle q^2} \right\} \right)^k \right\} }{\left( 1+\E\left\{ \frac{\displaystyle z^2}{\displaystyle q^2} \right\} \right)^k}\right] \,.
\end{align}
We are left with the derivation of $ \E\left\{ z^2 / q^2 \right\}$. We start noting that, from Lemma~\ref{lemma:ratio}, we have 
\begin{align}\label{e:ratio_of_squares}
\lefteqn{\E\left\{ \frac{\displaystyle z^2}{\displaystyle q^2} \right\} 
= \E \left\{ z^2 \right\} \E\left\{ \frac{1}{q^2} \right\} = \frac{\E \{z^2\}}{\E \{q^2\}} } & \notag \\
& \quad \times \left[ 1+\sum_{k=1}^\infty \left( -1 \right)^k \frac{\displaystyle \E \left\{ \left( q^2-\E\left\{ q^2 \right\} \right)^k \right\}}{\left(\E\left\{ q^2 \right\}\right)^k} \right] \,.
\end{align}
The variable $q^2$ has a non-central chi-squared distribution of 1 degree of freedom, thus $ \E\left\{ q^2 \right\} = \sigma_q^2 + \mu_q^2$,  and
\begin{align}
\E\left\{ \left( q^2 - \E\left\{q^2\right\} \right)^k \right\}
= & \sigma_q^{2k}\E\left\{ \left( \frac{q^2}{\sigma_q^2} - \E\left\{\frac{q^2}{\sigma_q^2}\right\} \right)^k \right\} \notag \\
= & \sigma_q^{2k} \mu_k \,.
\end{align}
This implies that 
\begin{align*}
\MoveEqLeft \lim_{k \rightarrow \infty } \frac{\displaystyle \E \left\{ \left( q^2-\E\left\{ q^2 \right\} \right)^k \right\}}{\left(\E\left\{ q^2 \right\}\right)^k} 
= \lim_{k \rightarrow \infty }\frac{\sigma_q^{2k} \mu_k}{(\mu^2_q+\sigma_q^{2})^k} \notag \\
& \quad = \lim_{k \rightarrow \infty }\frac{\mu_k}{\left(\frac{\mu^2_q}{\sigma_q^{2}}+1\right)^k} = 0 \,,
\end{align*}
since $\mu_k$ is bounded and the denominator grows unbounded, and condition~\eqref{conv_condition} holds, which is needed for the convergence of the sum in Eq.~\eqref{e:ratio_of_squares}.
Therefore, Eq.~\eqref{e:ratio_of_squares} becomes
\begin{align} \label{e:ratio_of_squares_closed_form}
\E\left\{ \frac{\displaystyle z^2}{\displaystyle q^2} \right\} = 
\frac{\displaystyle \mu_z^2 + \sigma_z^2}{\displaystyle \mu_q^2 + \sigma_q^2}
\left[ 1 + \sum_{k=1}^\infty \left( -1 \right)^k 
\frac{\sigma_q^{2k} \mu_k}{\left( \mu_q^2 + \sigma_q^2 \right)^k}
\right] \,.
\end{align}

We are left with the computation of the expectation in the numerator of the summation in~\eqref{e:closed_form_diag_init}. 
This reduces to the k-th central moment of a doubly noncentral F-distribution. With this goal in mind, we first define $\chi_z \triangleq z^2 / \sigma_z^2$ and $\chi_q \triangleq q^2 / \sigma_q^2$. We note that $\chi_z$ and $\chi_q$ are distributed according to a noncentral chi-squared distribution with 1 degree of freedom.
Therefore $ z^2/q^2 = \chi_z/\chi_q$, since $\sigma_z^2=\sigma_q^2$, where the random variable $\chi_z / \chi_q$ is distributed according to a doubly noncentral F-distribution.
Then, we have that
\begin{align} \label{e:F_distribution}
\E\left\{ \left(\frac{\chi_z}{\chi_q} - \E\left\{\frac{\chi_z}{\chi_q}\right\} \right)^k \right\}
=  \mu_{F,k} \,.
\end{align}

The proposition follows by substituting \eqref{e:ratio_of_squares_closed_form} and \eqref{e:F_distribution} in \eqref{e:closed_form_diag_init}, and noting that as required by~\eqref{conv_condition},   
\begin{align*}
\lim_{k\rightarrow \infty}\frac{\displaystyle  \E \left\{ \left( \frac{\displaystyle z^2}{\displaystyle q^2} - \E\left\{ \frac{\displaystyle z^2}{\displaystyle q^2} \right\} \right)^k \right\} }{{\left( 1+\E\left\{ \frac{\displaystyle z^2}{\displaystyle q^2} \right\} \right)^k}} =0\,,
\end{align*}
since the numerator is given by~\eqref{e:F_distribution} and is bounded, whereas the denominator diverges because is the power of a number greater than 1 due to that the expectation in the denominator is positive.

\section{Proof of Proposition \ref{prop:diag}}
\label{appendix_prop_diag}
The upper bound follows immediately from the fact that
$q^2 / \left(q^2+z^2\right) \leq 1, \, \forall q \in \R, \forall z \in \R$.
Regarding the lower bound, by noting that $\ln\left(\cdot\right)$ is a
concave function and applying Jensen's inequality, we obtain that
\begin{align} \label{e:diag1}
e^{\E\left\{ \ln\left(q^2\right) \right\} - \ln{\left( \E\left\{ q^2
    \right\} + \E\left\{ z^2 \right\} \right)}} \leq \E
\left\{\frac{q^2}{q^2+z^2}\right\} \,.
\end{align}
Moreover, $ \E\left\{ \ln\left(q^2\right) \right\} = - \gamma_e -
\ln{\left(2\right)} - 2\ln{\left(\sigma_q\right)} - M\left( 0, \,
\frac{1}{2}, \, -\mu_q^2/2\sigma_q^2 \right) $, whereas the proposition follows by substituting the previous equation in
\eqref{e:diag1}.

\section{Proof of Proposition \ref{prop:ublbpd}}
\label{appendix_prop_ublbpd}
We begin by observing that $\forall i,j$ the two matrices $\J^{(ub)}$
and $\J^{(lb)}$ satisfy $\J^{(lb)}_{ij} \leq \J_{ij} \leq
\J^{(ub)}_{ij}$. Consider the element-wise upper bound
$\J^{(ub)}_{ij}$. It follows that there exists a matrix $\G$
such that $\J_{ij} + \G_{ij} \geq \J^{(ub)}_{ij} \forall i,j$ and $\J
+ \G = \J^{(ub, G)}$, where $\G$ is symmetric and diagonally dominant
with real and non-negative elements. From the Gershgorin
circle theorem, \cite{Horn}, it follows that all its eigenvalues are
non-negative, i.e., $\lambda(\G) \geq 0$. Thus, $\lambda\left(\J^{(ub,
  G)}\right) = \lambda\left(\J\right) + \lambda\left(\G\right)$
because both $\J$ and $\G$ are Hermitian.  Therefore, since $\J$ is
positive semidefinite and $\lambda\left(\J^{(ub, G)}\right) >
\lambda\left(\J\right)$, $\J \preceq \J^{(ub, G)}$ and $\mathbf{0}
\preceq \J^{(ub, G)}$. The proof is concluded by noting that the same
steps may be employed for the lower bound $\J^{(lb, G)}$.


%

\ifCLASSOPTIONcaptionsoff
  \newpage
\fi



\bibliographystyle{IEEEtran}
\bibliography{IEEEabrv,coop_loc}
\enlargethispage{-5in}

\end{document}